\documentclass[english,aps,pra, twocolumn, footinbib, showpacs, showkeys, superscriptaddress,eqsecnum,longbibliography]{revtex4-1}
\usepackage[T1]{fontenc}
\usepackage{color}
\usepackage{babel}
\usepackage{bm}
\usepackage{amsmath}
\usepackage{amssymb}
\usepackage{graphicx}
\usepackage[unicode=true,
 bookmarks=true,bookmarksnumbered=false,bookmarksopen=false,
 breaklinks=true,pdfborder={0 0 1},backref=false,colorlinks=true]
 {hyperref}
\hypersetup{
 pdfauthor={Konschelle, Tokatly, Bergeret},
 citecolor=blue,linkcolor=magenta,urlcolor=blue}
\usepackage{breakurl}

\makeatletter
\usepackage{eucal}

\DeclareMathOperator{\Tr}{Tr}
\DeclareMathOperator{\sgn}{sgn}
\DeclareMathOperator{\Pexp}{Pexp}

\makeatother

\begin{document}

\title{Ballistic Josephson junctions in the presence of generic spin dependent
fields}

\author{Fran\c{c}ois Konschelle}

\affiliation{Centro de F\'{\i}sica de Materiales (CFM-MPC), Centro Mixto CSIC-UPV/EHU,
Manuel de Lardizabal 5, E-20018 San Sebasti\'{a}n, Spain}

\affiliation{JARA-Institute for Quantum Information, RWTH Aachen University, D-52074
Aachen, Germany}

\author{Ilya V. Tokatly}

\affiliation{Nano-Bio Spectroscopy group, Dpto. F\'{\i}sica de Materiales, Universidad
del Pa\'{\i}s Vasco, Av. Tolosa 72, E-20018 San Sebasti\'{a}n, Spain }

\affiliation{IKERBASQUE, Basque Foundation for Science, E-48011 Bilbao, Spain}

\author{F. Sebastian Bergeret}

\affiliation{Centro de F\'{\i}sica de Materiales (CFM-MPC), Centro Mixto CSIC-UPV/EHU,
Manuel de Lardizabal 5, E-20018 San Sebasti\'{a}n, Spain}

\affiliation{Donostia International Physics Center (DIPC), Manuel de Lardizabal
4, E-20018 San Sebasti\'{a}n, Spain}
\begin{abstract}
Ballistic Josephson junctions are studied in the presence of a spin-splitting
field and spin-orbit coupling. A generic expression for the quasi-classical
Green's function is obtained and with its help we analyze several
aspects of the proximity effect between a spin-textured normal metal
(N) and singlet superconductors (S). In particular, we show that the
density of states may show a zero-energy peak which is a generic consequence
of the spin-dependent couplings in heterostructures. In addition we
also obtain the spin current and the induced magnetic moment in a
SNS structure and discuss possible coherent manipulation of the magnetization
which results from the coupling between the superconducting phase
and the spin degree of freedom. Our theory predicts a spin accumulation
at the S/N interfaces, and transverse spin currents flowing perpendicular
to the junction interfaces. Some of these findings can be understood
in the light of a non-Abelian electrostatics.
\end{abstract}

\pacs{74.50.+r Tunneling phenomena; Josephson effects - 74.78.Na Mesoscopic
and nanoscale systems - 85.25.Cp Josephson devices - 72.25.-b Spin
polarized transport}

\keywords{Josephson junction ; superconducting heterostructures ; current-phase
relation ; spin current ; spin polarization ; spin capacitor ; transverse
spin current ; spin-dependent density of states ; ballistic mesoscopic
system ; spin-orbit ; magnetic texture ; zero-energy peak ; transport
equation ; gauge-covariant quasi-classic Green functions}

\date{\today}

\maketitle

\section{Introduction}

There are great hopes that a low dissipative spintronics might emerge
from the combination of superconducting and magnetic materials \citep{Eschrig2011,Linder2015,Eschrig2015a}.
In addition, the intrinsic coherence associated with superconducting
transport might well lead to important discoveries, ranging from technological
applications in the fields of quantum circuitry \citep{Xiang2013}
and quantum computation \citep{Nayak2008,Alicea2014}, to fundamental
perspectives in the understanding of the interactions between superconductivity
and magnetism \citep{Casalbuoni2004,buzdin.2005_RMP,Bergeret2005}.

Superconducting spintronics applications mainly lie in the possibility
to generate spin-polarized Cooper pairs, the so-called triplet correlations,
in heterostructures combining ferromagnets (F) and superconductors
(S) \citep{Bergeret2005}, which have been explored intensively in
the last years \citep{Robinson2010,Khaire2010,Anwar2010,Anwar2012a,Visani2012,Khaydukov2014,Singh2015,Kalcheim2015}.

Also promising for coherent spintronics applications are recent proposals
 for coupling of charge and spin degrees of freedom by combining superconductors
and materials with strong spin-orbit (SO) interactions \citep{Wakamura2014,Wakamura2015}.
Of particular interest are the possibilities to generate phase dependent
spin currents \citep{Malshukov2008,Malshukov2010,Alidoust2015a,Alidoust2015,Konschelle2015},
to manipulate the magnetization dynamics coherently \citep{Konschelle2009,Kulagina2014},
and to exploit magneto-electric effects in S/N/S structures \citep{Konschelle2015,Bergeret2014a}
for creating supercurrents polarizing the junctions. 

A quantitative description of spin-dependent transport in superconducting
systems necessarily implies an accurate description of the proximity
effect between the magnetic and superconducting elements \citep{buzdin.2005_RMP,Bergeret2005}.
This is accounted for in the so-called quasi-classical formalism,
based on the Eilenberger equation \citep{Eilenberger1968,Larkin1969}.

The quasi-classical approach has been recently generalized to describe
the coupling between the spin and charge degrees of freedom in superconducting
heterostructures with intrinsic SO coupling \citep{Bergeret2014,Konschelle2014,Konschelle2015,Bergeret2013}.
In particular, the dominant phenomenologies of a ballistic S/N/S Josephson
junction with a generic intrinsic spin dependent field are described
by only two parameters: a phase $\Phi$ and a unit vector $\boldsymbol{\mathfrak{n}}$
\citep{Konschelle2016}. Within the quasi-classical approach the unit
vector $\boldsymbol{\mathfrak{n}}$ describes the local spin quantization
axis about which the classical spin precesses at a constant latitude
while propagating through the junction along the Andreev-modes trajectories,
whereas the angle $\Phi$ measures the mismatch of the precession
angle after a quasiparticle completes the semiclassical loop (Andreev
loop) in the normal metal. 

In the present work we use the formalism developed in Ref. \citep{Konschelle2016}
to study the spin and charge observables of a ballistic Josephson
S/N/S junction with arbitrary spin dependent fields. From the general
Eilenberger equation, that takes into account charge and spin degrees
of freedom on equal footing (section \ref{sec:Equation-of-motion}),
we obtain a generic expressions for the quasi-classic Green's function
all over the coherent structure (section \ref{sec:General-Solution}).
From its knowledge we analyze spin and charge current-phase relations
in sections \ref{sec:Density-of-states}-\ref{sec:Spin-effects},
provide several examples of non-trivial $\Phi$ and $\boldsymbol{\mathfrak{n}}$
quantities, and discuss their connection with charge and spin observables. 

We show that the phase $\Phi$ completely encrypts the effect of the
spin fields on the charge observables, namely the charge current-phase
relation and the density of state for a generic magnetic interaction
\ref{sec:Density-of-states}. In particular we demonstrate the presence
of a peak at zero-energy in the density of state, which is a generic
consequence of a non trivial magnetic angle $\Phi$. As examples we
analyze the current-phase relation for an anti-ferromagnetic ordering,
a monodomain ferromagnet with spin-orbit interaction (section \ref{sec:SFS}),
and a Bloch domain-wall (section \ref{sec:Bloch-domain-wall}). 

Spin observables not only depends on the angle $\Phi$ but also on
the vector $\boldsymbol{\mathfrak{n}}$, as shown in section \ref{sec:Spin-effects}.
We discuss different cases when either Dresselhaus and/or Rashba spin-orbit
interactions are present in the junction. We show that the spin current
disappears when either the exchange or the spin-orbit interaction
vanishes, whereas the spin polarization survives the absence of a
spin-orbit coupling. In this later case we predict a reversal of this
extra contribution with respect to the temperature and length for
specific values of the superconducting phase difference $\varphi$
and magnetic angle $\Phi$ in section \ref{sec:Density-of-states}. 

We finally demonstrate that all spin observables can be expressed
in terms of the SU(2) electric field (sections VI and VII) and its
covariant derivatives. In particular the spin polarization and spin
currents obey a non-Abelian generalization of the Maxwell equations
in electrostatics. In leading order of the spin fields we recognize
two intriguing effects, namely the accumulation of the spin polarization
at the interfaces between the spin textured region and the superconducting
banks, which we call the \textit{spin capacitor effect,} and the generation
of a transverse spin current along the superconducting interfaces,
due to some \textit{displacement spin currents}. All our predictions
can be measured using state-of-the-art experimental techniques and
can be seen as precursors of the actively searched topological effects
in superconducting heterostructures.

\section{The Model\label{sec:Equation-of-motion}}

In this study we consider a Josephson junction made of two $s$-wave
superconductors S connected by a normal region N of length $L$. The
phase difference between the S electrodes is $\varphi$. The normal
metal hosts spin dependent fields, both spin-orbit or spin-splitting
ones.

The total Hamiltonian of the system reads 
\begin{equation}
H=H_{0}+W_{\text{BCS}}
\end{equation}
with 
\begin{equation}
H_{0}=\int d\boldsymbol{r}\left[\Psi^{\dagger}\left(\dfrac{\boldsymbol{p}^{2}}{2m}-\mu-\boldsymbol{h\cdot\sigma}-\dfrac{A_{i}^{a}p_{i}\sigma^{a}}{2m}\right)\Psi\right]\label{eq:H-0}
\end{equation}
being the one body Hamiltonian. Here $\Psi(\bm{r})$ and $\Psi^{\dagger}(\bm{r})$
are the spinor field operators, $\mu$ is the chemical potential,
$m$ is the effective mass, $\boldsymbol{h\cdot\sigma}$ is the spin-splitting
(exchange or Zeeman) field, and $A_{i}^{a}p_{i}\sigma^{a}/2$ describes
the spin-orbit coupling which is assumed to be linear in momentum
$\bm{p}$. Throughout this paper the sum over repeated indices is
implied and the lower (upper) indices denote spatial (spin) coordinates.
The matrices $\sigma^{1,2,3}$ are Pauli matrices spanning the spin
algebra. A particular case of $A_{x}^{y}=-A_{y}^{x}=\alpha$, in Eq.\textasciitilde{}\eqref{eq:H-0}
corresponds to the Rashba spin-orbit coupling, whereas $A_{x}^{x}=-A_{y}^{y}=\beta$
is the Dresselhaus spin-orbit coupling. 

The superconducting correlations are described by the usual BCS interaction
term $W_{{\rm BCS}}$ in the S electrodes, 
\begin{equation}
W_{\text{BCS}}=\int d\boldsymbol{r}\left[\dfrac{V}{2}\left(\Psi\mathbf{i}\sigma^{2}\Psi\right)^{\dagger}\left(\Psi\mathbf{i}\sigma^{2}\Psi\right)\right]\;,\label{eq:H-BCS}
\end{equation}
which we treat in the standard BCS mean field approximation. Notice
that we assume that the spin-dependent fields are zero in the S electrodes.

As long as the spin-orbit coupling is linear in momentum one can write
the one body Hamiltonian \eqref{eq:H-0} as follows 
\begin{equation}
H_{0}=\int d\boldsymbol{r}\left[\Psi^{\dagger}\left(\dfrac{\left(\boldsymbol{p}-\boldsymbol{A}\right)^{2}}{2m}-\tilde{\mu}-A_{0}\right)\Psi\right]\quad.\label{eq:H-0-A}
\end{equation}
Passing from \eqref{eq:H-0} to \eqref{eq:H-0-A} imposes shifting
the chemical potential $\mu\rightarrow\tilde{\mu}=\mu-\left(A_{i}^{a}\right)^{2}/8m$,
without any physical consequence. Now the $2\text{\ensuremath{\times}\ 2}$
vector-valued matrix $\boldsymbol{A}\equiv A_{i}^{a}\sigma^{a}/2$
can be interpreted as a non-Abelian SU(2) gauge potential in the space
sector, and the quantity $A_{0}\equiv\boldsymbol{h\cdot\sigma}$ can
be viewed as a gauge potential in the time sector \citep{Frohlich1992,Frohlich1993,Berche2013,Tokatly2008b}.
The understanding of the spin-splitting and spin-orbit effects in
terms of the gauge potential is appealing, since it allows a straightforward
perturbation scheme to be implemented, with the strong requirement
that any order must be gauge covariant. Then the strategy is to promote
the model to be gauge invariant (note that $W_{\text{BCS}}$ is invariant
with respect to any spin rotation since it describes $s$-wave pairing
and hence it is already $\text{SU}\left(2\right)$ gauge invariant),
and to obtain a set of gauge invariant observables that one can calculate
with any accuracy using a covariant perturbation scheme.

To describe superconducting heterostructures it is convenient to employ
the so-called quasi-classic method valid when any characteristic length
scale $\xi$ involved in the problem is much larger than the Fermi
wavelength $1/p_{F}$ \citep{serene_rainer.1983,Rammer1986,Langenberg1986,Wilhelm1999,b.kopnin}.
In the lowest order in $\xi p_{F}$ the resulting kinetic-like equation
is the so called Eilenberger equation for the quasi-classical Green's
function $\check{g}(\bm{r})$. In the presence of non-Abelian gauge-potentials
the Eilenberger equation reads \citep{Konschelle2015,Bergeret2014,Konschelle2014}
(we set $\hbar=1$)
\begin{equation}
\mathbf{i}\left(\boldsymbol{v\cdot\mathfrak{D}}\right)\check{g}+\left[\tau_{3}\left(\omega+A_{0}\right)+\check{\Delta},\check{g}\right]=0\;,\label{eq:Eilenberger}
\end{equation}
where $\mathfrak{D}_{i}\check{g}=\partial_{i}\check{g}-\mathbf{i}\left[A_{i},\check{g}\right]$
is the covariant derivative and 
\begin{equation}
\check{\Delta}\left(x\right)=\left(\begin{array}{cc}
0 & \Delta\left(x\right)\\
-\Delta^{\ast}\left(x\right) & 0
\end{array}\right)
\end{equation}
as the mean-field anomalous self energy in the Nambu space. The superconducting
order parameter $\Delta\left(x\right)$ is proportional to the unit
matrix in the spin space. In the equilibrium situation considered
here, the quasi-classical Green's function depends on the direction
of the Fermi velocity $\bm{v}$ ($\left|\bm{v}\right|=v_{F}$) and
on the frequency $\omega$, and has the following general form 
\begin{equation}
\check{g}\left(x\right)=\left(\begin{array}{cc}
g & f\\
\mathcal{T}f\mathcal{T}^{-1} & -\mathcal{T}g\mathcal{T}^{-1}
\end{array}\right)\;,\label{eq:g-parameterisation}
\end{equation}
where $g$ and $f$ are matrices in the spin space, and $\mathcal{T}=\mathcal{K}\mathbf{i}\sigma_{2}$
represents the time-reversal operation with $\mathcal{K}$ being the
operation of complex-conjugation supplemented with reversal of $\bm{v}$
and the real part of $\omega$, so that $\mathcal{T}^{2}=-1$. It
is worth noting that the structure of Eq.\eqref{eq:g-parameterisation}
verifies a particle-hole symmetry $\left\{ \mathcal{P},\check{g}\right\} =0$
with $\mathcal{P}=\mathcal{K}\tau_{2}\sigma_{2}$, where the $\tau$'s
are Pauli matrices in the particle-hole (or Nambu) space. 

One could include in Eq.\eqref{eq:Eilenberger} a collision term due
to scattering at impurities, however here we only consider the pure
ballistic limit. In addition, one can also consider higher order terms
in $\xi p_{F}$ and include in Eq. \eqref{eq:Eilenberger} the effect
of a non-Abelian Lorentz force due to the SU(2) magnetic field (see
\citep{Konschelle2015,Bergeret2014,Konschelle2014}). These terms
are responsible for the spin Hall effect and the anomalous Josephson
phase $\varphi_{0}$ \citep{Bergeret2014a,Konschelle2015}. Below
we disregard these effects and study the physics governed by the Eilenberger
equation at the level of the leading quasi-classical order. In this
approximation the spin-orbit field $\bm{A}$ leads to the spin precession
via the commutator part of the covariant derivative in Eq.\eqref{eq:Eilenberger}.~

Our goal is to calculate the physical observables, namely, the charge
current 
\begin{equation}
j_{i}\left(\boldsymbol{x}\right)=-\mathbf{i}\dfrac{\pi}{2}eN_{0}k_{B}T\sum_{n=-\infty}^{\infty}\Tr\left\langle v_{i}\tau_{3}\check{g}\left(\mathbf{i}\omega_{n}\right)\right\rangle \;,\label{eq:current-clean}
\end{equation}
the spin current
\begin{equation}
\mathfrak{J}_{i}^{a}\left(\boldsymbol{x}\right)=-\mathbf{i}\dfrac{\pi}{2}N_{0}k_{B}T\sum_{n=-\infty}^{\infty}\Tr\left\langle v_{i}\sigma^{a}\check{g}\left(\mathbf{i}\omega_{n}\right)\right\rangle \;,\label{eq:current-spin}
\end{equation}
and the electronic spin density 
\begin{equation}
S^{a}\left(\boldsymbol{x}\right)=-\mathbf{i}\dfrac{\pi}{2}N_{0}k_{B}T\sum_{n=-\infty}^{\infty}\Tr\left\langle \sigma^{a}\tau_{3}\check{g}\left(\mathbf{i}\omega_{n}\right)\right\rangle \;.\label{eq:density-spin}
\end{equation}
in the junction. Here $\omega_{n}=2\pi k_{B}T\left(n+1/2\right)$
are the Matsubara frequencies and $\left\langle \cdots\right\rangle $
represents the ~angular averaging over the Fermi surface and $N_{0}$
is the density of states at the Fermi level, in any dimension. It
is important to emphasize that $S$ calculated from the quasi-classical
Green's function denotes the change of the spin polarization due to
the electrons at the Fermi level and not the total magnetic moment.
The total magnetization is obtained by adding to $S$ the Pauli paramagnetic
contribution $\sim N_{0}h$ \citep{Bergeret2004}. 

As defined in \eqref{eq:current-clean}, the charge current is conserved
$\boldsymbol{\nabla\cdot j}=0$ whereas the spin observables are covariantly
conserved $\mathfrak{D}_{t}S+\mathfrak{D}_{i}\mathfrak{J}_{i}=0$,
provided the gap-parameter is obtained self-consistently $\Delta=-\mathbf{i}\pi\hbar VN_{0}k_{B}T\sum_{n}\left\langle f\right\rangle $.
In the following and for simplicity we disregard the difficulty of
dealing with the self-consistent condition and assume that the weak-link
does not alter $\Delta$. This assumption works well for short junctions,
however for long junctions, the self-consistency condition should
not be ignored, since phase-slips may arise \citep{Martin-Rodero1993,Sols1994,LevyYeyati1995,Riedel1996}.
In such a case we assume that the superconducting gap in the electrodes
is induced by the proximity effect from a large superconductor. This
describes for example a lateral junction made by a 2D electron gas
with 3D superconducting electrodes deposited on top of the 2D-system.
 In such a case the self-consistency can be avoided and the rigidity
of $\Delta$ is justified.

\section{General Solution Of the Eilenberger Equation\label{sec:General-Solution}}

We now solve the Eilenberger equation \eqref{eq:Eilenberger} for
a Josephson junction consisting of a normal metal bridge of length
$L$ with magnetic interaction sandwiched between two superconducting
electrodes phase-shifted by $\varphi$. We assume that the dimensions
perpendicular to the junction axis are much larger than $L$, then
the problem is quasi-1D: $d\check{g}=\left(\boldsymbol{v\cdot\nabla}\right)\check{g}ds$,
which transforms \eqref{eq:Eilenberger} to the simple rotation equation
\begin{equation}
\mathbf{i}\dfrac{d\check{g}}{ds}+\left[\tau_{3}\omega+\tau_{3}A_{0}+\boldsymbol{v\cdot A}+\check{\Delta},\check{g}\left(s\right)\right]=0\label{eq:transport-Heisenberg}
\end{equation}
where $\check{g}\left(s\right)$ is a short-hand notation for $\check{g}\left(x\left(s\right),y\left(s\right),\cdots\right)$.
For any Fermi surface, one can choose for instance $s=x/v_{x}$, with
the $x$-axis along the junction. Using the Ansatz \citep{Schopohl1995,Schopohl1998}
\begin{equation}
\check{g}\left(s\right)=\check{u}\left(s,s_{0}\right)\check{g}\left(s_{0}\right)\check{u}\left(s_{0},s\right)+\check{g}_{\infty}\label{eq:Ansatz}
\end{equation}
with $\check{g}\left(s_{0}\right)$ and $\check{g}_{\infty}$ some
constant matrices, the transport equation \eqref{eq:transport-Heisenberg}
reduces to the equation for the propagator $\check{u}\left(s,s_{0}\right)$
\begin{equation}
\mathbf{i}\dfrac{d\check{u}\left(s,s_{0}\right)}{ds}+\left(\tau_{3}\omega+\tau_{3}A_{0}+\boldsymbol{v\cdot A}+\check{\Delta}\right)\check{u}\left(s,s_{0}\right)=0\label{eq:transport-Schrodinger}
\end{equation}
with boundary condition $\check{u}\left(s_{0},s_{0}\right)=1$ and
$\check{u}\left(s_{2},s_{0}\right)=\check{u}\left(s_{2},s_{1}\right)\check{u}\left(s_{1},s_{0}\right)$,
in addition to the relation 
\begin{equation}
\left[\tau_{3}\omega+\tau_{3}A_{0}+\boldsymbol{v\cdot A}+\check{\Delta},\check{g}_{\infty}\right]=0
\end{equation}
Note that this last commutator-equation can only be verified when
$A_{0}$, $\boldsymbol{A}$ and $\Delta$ are $s$-independent, in
order for $\check{g}_{\infty}$ to be $s$-independent. So this equation
must be verified only for large $s$, or equivalently for bulk systems.

In the following we assume that in the superconducting electrodes
$A_{0}=\boldsymbol{v\cdot A}=0$ and that $\Delta$ is constant, while
in the normal region $\Delta=0$. In this case Eq.\eqref{eq:transport-Schrodinger}
can be easily integrated.

If the superconducting electrodes are located at $s\leq s_{L}$ and
$s\geq s_{R}$, and the phase difference between them is $\varphi$,
the general solution in superconducting regions can be written in
the form
\begin{align}
\check{g}\left(s\leq s_{L}\right) & =e^{\mathbf{i}\tau_{3}\frac{\varphi}{4}}\mathbf{S}_{L}\left[g_{1}\tau_{+}-\tau_{3}\right]\mathbf{S}_{L}^{-1}e^{-\mathbf{i}\tau_{3}\frac{\varphi}{4}}\nonumber \\
\check{g}\left(s\geq s_{R}\right) & =e^{-\mathbf{i}\tau_{3}\frac{\varphi}{4}}\mathbf{S}_{R}\left[g_{2}\tau_{-}-\tau_{3}\right]\mathbf{S}_{R}^{-1}e^{\mathbf{i}\tau_{3}\frac{\varphi}{4}}\label{eq:sol-supra}
\end{align}
with 
\begin{equation}
\mathbf{S}_{L,R}\left(s\right)=\dfrac{e^{\mathbf{i}\eta/2}-\mathbf{i}\tau_{1}e^{-\mathbf{i}\eta/2}}{\sqrt{2\cos\eta}}e^{\Delta\left(s-s_{L,R}\right)\tau_{3}\cos\eta}
\end{equation}
and $\sin\eta=\omega/\Delta$. $g_{1,2}$ are some constant matrices
in the spin-space to be determined by boundary conditions. The matrices
$\tau_{\pm}$ select the physically acceptable evanescent waves in
\eqref{eq:sol-supra}, whereas the matrix $\mathbf{S}_{L}\left(-\tau_{3}\right)\mathbf{S}_{L}^{-1}=\mathbf{S}_{R}\left(-\tau_{3}\right)\mathbf{S}_{R}^{-1}$
represents the bulk solution of the superconductors, far away from
the interfaces where the evanescent waves $e^{\pm\Delta\left(s-s_{L,R}\right)\cos\eta}$
disappear. We note that $\check{g}^{2}=1$ at any $s$ in \eqref{eq:sol-supra}.

The expression \eqref{eq:sol-supra} is for the positive velocity
only. The negative velocity counterpart is found by the substitution
$e^{\Delta\left(s-s_{L,R}\right)\tau_{3}\cos\eta}\rightarrow e^{-\Delta\left(s-s_{L,R}\right)\tau_{3}\cos\eta}$
and $\tau_{+}\leftrightarrow\tau_{-}$ in \eqref{eq:sol-supra} in
order to select the evanescent waves decaying from the interfaces
towards the bulk superconductors.

The solution in the normal region is simpler. Since there is no gap
there Eq.\eqref{eq:transport-Schrodinger} takes the form 
\begin{equation}
\left[\mathbf{i}\dfrac{d}{ds}+\tau_{3}\omega+\tau_{3}A_{0}+\boldsymbol{v\cdot A}\right]\check{u}_{N}\left(s,s_{0}\right)=0\;,\label{eq:schrodinger-uN}
\end{equation}
and can be integrated as
\begin{equation}
\check{u}_{N}=e^{\mathbf{i}\tau_{3}\omega s}\left(\begin{array}{cc}
u & 0\\
0 & \bar{u}
\end{array}\right)\;;\;\bar{u}=\mathcal{T}u\mathcal{T}^{-1}\label{eq:u-parameterisation}
\end{equation}
because $A_{0}$ and $\boldsymbol{v\cdot A}$ are matrices in the
spin-space only, and thus they commute with $\tau_{3}$. The remaining
spin propagator $u\left(s,s_{0}\right)$ satisfies the equation
\begin{equation}
\left[\mathbf{i}\dfrac{d}{ds}+A_{0}+\boldsymbol{v\cdot A}\right]u\left(s,s_{0}\right)=0\label{eq:u}
\end{equation}
where the gauge potentials $A_{0}$ and $\bm{A}$ as well as the velocity
$\bm{v}$ can be $s$-dependent. One of the interests of this study
is to establish a generic current-phase relations without any specific
assumption about the configuration of the spin-dependent fields and
the shape of the Fermi surface. Equation \eqref{eq:u} can be solved
by applying any usual perturbation scheme, see \textit{e.g.} \citep{Blanes2009}.
In the most general case, $u\left(s_{2},s_{1}\right)$ verifies $u\left(s,s\right)=1$,
$u\left(s_{1},s_{2}\right)=u\left(s_{2},s_{1}\right)^{\dagger}$,
and can be represented as follows 
\begin{equation}
u\left(s_{2},s_{1}\right)=\Pexp\left\{ \mathbf{i}\int_{s_{1}}^{s_{2}}ds\left[A_{0}\left(s\right)+\boldsymbol{v\cdot A}\left(s\right)\right]\right\} \label{eq:path-integral}
\end{equation}
where $\Pexp$ stands for the path-ordered exponential along the path
connecting points $s_{1}$ to $s_{2}$. The operator $u\left(s_{2},s_{1}\right)$
($\bar{u}\left(s_{2},s_{1}\right)$) propagates the electron (hole)
component of the full Green's function $\check{g}_{N}\left(s_{2}\right)=\check{u}_{N}\left(s_{2},s_{1}\right)\check{g}_{N}\left(s_{1}\right)\check{u}_{N}\left(s_{1},s_{2}\right)$
from the point $s_{1}$ to the point $s_{2}$, both inside the normal
region.

We now proceed to construct the quasi-classical Green's function in
the whole space from $s\rightarrow-\infty$ to $s\rightarrow+\infty$
by matching the solutions \eqref{eq:sol-supra} in the S-electrodes
with the solution in the N-region,
\begin{equation}
\check{g}\left(s_{L}\leq s\leq s_{R}\right)=\check{u}_{N}\left(s,s_{0}\right)\check{g}\left(s_{0}\right)\check{u}_{N}\left(s_{0},s\right)\quad,\label{eq:g_in_N}
\end{equation}
where $\check{u}_{N}$ is defined in \eqref{eq:u-parameterisation}
and $s_{0}\in\left[s_{L},s_{R}\right]$ is an arbitrary origin of
coordinates. 

Assuming perfectly transparent interfaces at $s_{L,R}=s\left(x=\mp L/2\right)$,
we impose the continuity of the matrix $\check{g}$ (cf. Eqs. \eqref{eq:g_in_N}
and \eqref{eq:sol-supra}) 
\begin{multline}
e^{\mathbf{i}\tau_{3}\frac{\varphi}{4}}\mathbf{S}_{L}\left(s_{L}\right)\left[g_{1}\tau_{+}-\tau_{3}\right]\mathbf{S}_{L}^{-1}\left(s_{L}\right)e^{-\mathbf{i}\tau_{3}\frac{\varphi}{4}}\\
=\check{u}_{N}\left(s_{L},s_{0}\right)\check{g}\left(s_{0}\right)\check{u}_{N}\left(s_{0},s_{L}\right)
\end{multline}
\begin{multline}
e^{-\mathbf{i}\tau_{3}\frac{\varphi}{4}}\mathbf{S}_{R}\left(s_{R}\right)\left[g_{2}\tau_{-}-\tau_{3}\right]\mathbf{S}_{R}^{-1}\left(s_{R}\right)e^{\mathbf{i}\tau_{3}\frac{\varphi}{4}}\\
=\check{u}_{N}\left(s_{R},s_{0}\right)\check{g}\left(s_{0}\right)\check{u}_{N}\left(s_{0},s_{R}\right)\label{eq:boundary}
\end{multline}
These equations should uniquely determine the constant matrices $\check{g}\left(s_{0}\right)$,
$g_{1}$ and $g_{2}$. By eliminating $\check{g}\left(s_{0}\right)$
we get
\begin{equation}
\mathbf{Q}\left[g_{1}\tau_{+}-\tau_{3}\right]=\left[g_{2}\tau_{-}-\tau_{3}\right]\mathbf{Q}\label{eq:continuity-LR}
\end{equation}
\begin{multline}
\mathbf{Q}=\dfrac{e^{\mathbf{i}\eta/2}+\mathbf{i}\tau_{1}e^{-\mathbf{i}\eta/2}}{\sqrt{2\cos\eta}}e^{\mathbf{i}\tau_{3}\varphi/4}\times\\
\check{u}\left(s_{R},s_{L}\right)e^{\mathbf{i}\tau_{3}\varphi/4}\dfrac{e^{\mathbf{i}\eta/2}-\mathbf{i}\tau_{1}e^{-\mathbf{i}\eta/2}}{\sqrt{2\cos\eta}}\label{eq:Q}
\end{multline}
Equation \eqref{eq:continuity-LR}, being a matrix relation, corresponds
to a system of four equations for two $2\times2$ matrices, $g_{1}$
and $g_{2}$. However, there are only two linearly independent equations
that allows for uniquely determining $g_{1}$ and $g_{2}$. Once $g_{1}$
and $g_{2}$ are obtained, one calculates $\check{g}\left(s_{0}\right)$
from Eq.\eqref{eq:boundary}. Finally, using Eqs.\eqref{eq:sol-supra}
and \eqref{eq:schrodinger-uN}, and setting $s=x/v_{x}$, we find
$\check{g}(x)$ in the whole space (details of this calculation can
be found in Appendix \ref{app:Greens-functions}). 

To calculate the physical observables, Eqs.\eqref{eq:current-clean}-\eqref{eq:density-spin},
we need only the electron component $g(x)$ of full Green's function
in Eq.\eqref{eq:g-parameterisation}. In the N-region at $x_{L}<x<x_{R}$
the function $g\left(x\right)$ takes the form (see Appendix \ref{app:Greens-functions})

\begin{equation}
g\left(x\right)=\dfrac{W^{-1}\left(x\right)-W\left(x\right)-2\mathbf{i}\sin2\chi}{2\cos2\chi+\Tr\left\{ W\right\} }\label{eq:g0}
\end{equation}
\begin{equation}
\chi=\frac{\omega L}{\left|v_{x}\right|}+\sgn\left(v_{x}\right)\dfrac{\varphi}{2}+\arcsin\dfrac{\omega}{\Delta}\label{eq:chi}
\end{equation}
where we defined the operator
\begin{equation}
W\left(x\right)\equiv u\left(x,x_{L}\right)\bar{u}\left(x_{L},x_{R}\right)u\left(x_{R},x\right)\label{eq:WA}
\end{equation}
Physically the operator $W\left(x\right)$ describes the propagation
of an electron from the point $x$ to the right interface $x_{R}$,
where it is reflected as a hole towards the left interface $x_{L}$,
and finally converted back to an electron and returns to the initial
point $x$. Its inverse $W^{-1}\left(x\right)$ corresponds to the
opposite propagation: electron originally moving from $x$ to $x_{L}$
is converted to a hole there and goes back to $x_{R}$, where the
hole is converted to an electron again and return back to $x$ from
the opposite side. Therefore we see that $W$ and $W^{-1}$ represent
the two possible loops made by the particles trajectories in the junction,
according to the usual picture of Andreev modes \citep{kulik.1970}. 

It is interesting to note that the operator of Eq.\eqref{eq:WA} can
be interpreted as a kind of Wilson loop operator describing a SU(2)
holonomy in the effective $\mathbb{R}\times\mathbb{Z}_{2}$ parameter
space spanned by the coordinate $x$ and the electron-hole index.
The operator $W\left(x\right)$ Eq.\eqref{eq:WA} which mixes the
particle $u$ and anti-particle $\bar{u}$ propagators has been recently
introduced in the context of semiclassical quantization of spinning
Bogoliubov quasiparticles \citep{Konschelle2016}. Here we follow
the notation of Ref.\citep{Konschelle2016} and call $W(x)$ the \textit{Andreev-Wilson
loop operators}. 

\begin{figure}[b]
\includegraphics[width=0.95\columnwidth]{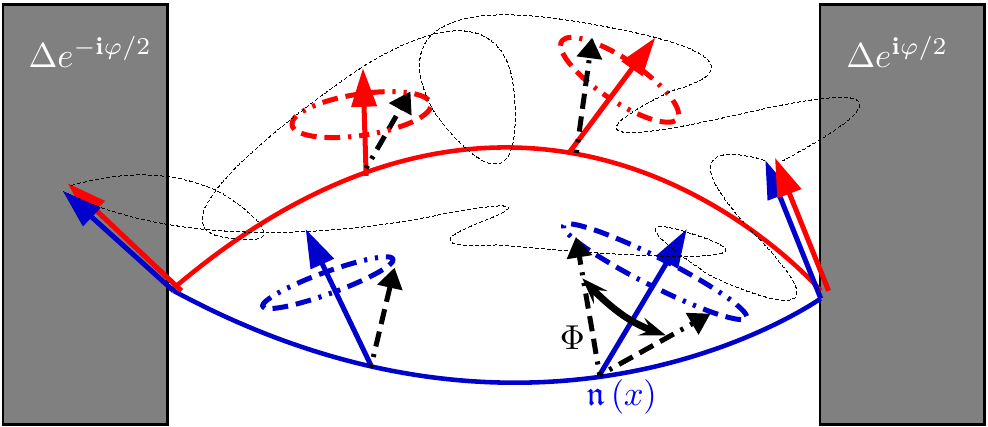}

\caption{\label{fig:AW-loop}Sketch of the Josephson junction, with $\varphi$
the phase-difference between the two superconducting electrodes (in
grey). The white part corresponds to the normal region, with a spin
texture giving rise to the Andreev-Wilson loop $W\left(x\right)$
at point $x$, see Eq.\eqref{eq:WA}. The electron spin precesses
at a constant latitude around the local vector $\boldsymbol{\mathfrak{n}}$
when travelling along the junction ($\boldsymbol{\mathfrak{n}}$ evolves
according to the equation \eqref{eq:n-eq}). Blue and red colors refer
to a precession axis for electrons and holes, respectively. After
completing the loop, the spin rotates an angle $\Phi$ between the
initial and final states.}
\end{figure}

Since $u$ is a SU(2) rotation matrix, so is the operator $W\left(x\right)$
obtained by a combination of rotations. One can thus parameterize
the Andreev-Wilson loop operator by a unit vector $\boldsymbol{\mathfrak{n}}$
and an angle $\Phi$ as follows 
\begin{equation}
W\left(x\right)=e^{\mathbf{i}\left(\boldsymbol{\mathfrak{n}\cdot\sigma}\right)\Phi}=\cos\Phi+\mathbf{i}\left(\boldsymbol{\mathfrak{n}\cdot\sigma}\right)\sin\Phi\,.\label{eq:WA-expansion}
\end{equation}
The parameters $\boldsymbol{\mathfrak{n}}$ and $\Phi$ are related
to the spin fields $A_{0}$ and $\bm{A}$ via the path-ordered representation,
Eq.\eqref{eq:path-integral}, of the propagators $u$ and $\bar{u}$
in Eq.\eqref{eq:WA}. Importantly, $\boldsymbol{\mathfrak{n}}$ and
$\Phi$ encode all physical effects of spin interactions (Zeeman and
spin orbit) as the latter enter the Green's function $g(x)$ only
via the Andreev-Wilson loop operator. 

By taking a trace of Eqs.\eqref{eq:WA} and \eqref{eq:WA-expansion}
we find that the quantity
\begin{align}
2\cos\Phi & =\Tr\left\{ W\left(x\right)\right\} =\Tr\left\{ \bar{u}\left(x_{L},x_{R}\right)u\left(x_{R},x_{L}\right)\right\} \label{eq:cosPhi}
\end{align}
appearing in the denominator of \eqref{eq:g0} is position independent.
Therefore the angle $\Phi$ is a $x$-independent global parameter
and only $\mathfrak{n}$ may depend on the position $x$ in \eqref{eq:WA-expansion}.
The unit vector $\boldsymbol{\mathfrak{n}}\left(x\right)$ determines
the local spin quantization axis. Semiclassically it can be viewed
as an axis about which the spin precesses at a constant latitude.
The angle $\Phi$ records the phase accumulated by the electron wave
function after one cycle along the Andreev loop (see
Fig. \ref{fig:AW-loop}). This is nothing but the holonomy associated
to the equation \eqref{eq:u} for $u$ along the Andreev loop.
The physical significance of the Andreev-Wilson loop operator is illustrated
on Fig.\ref{fig:AW-loop}. When an electron travels from the left
to the right electrode inside the normal region, its spin precesses
at a constant latitude about a position dependent axis $\boldsymbol{\mathfrak{n}}\left(x\right)$.
At the right interface it is converted into a hole according to the
scheme of Andreev reflection \citep{Andreev1964}. The spin of the
resulting hole, moving from the right to the left electrode, precesses
about the time-reversal conjugate of $\boldsymbol{\mathfrak{n}}$.
At the left interface another Andreev reflection transforms the hole
back to an electron-like particle with the spin precessing again about
the direction of $\boldsymbol{\mathfrak{n}}$. When returning to its
original position, the electron spinor accumulates an extra ``magnetic''
phase $\Phi$ according to the Andreev-Wilson loop operator $W\left(x\right)$
in \eqref{eq:WA-expansion}. The phase $\Phi$ is nothing but the
angle between the initial and final directions of the electron spin
\citep{Konschelle2016}.

By substituting $W(x)$ of Eq.\eqref{eq:WA-expansion} into Eq.\eqref{eq:g0}
and using the trigonometric identity
\begin{equation}
\tan\left(a+b\right)+\tan\left(a-b\right)=\dfrac{2\sin2a}{\cos2a+\cos2b}\label{eq:tangent}
\end{equation}
we find the following explicit representation for the quasi-classical
Green's function in the normal region $x_{L}\leq x\leq x_{R}$ 
\begin{multline}
g\left(x\right)=-\dfrac{\mathbf{i}}{2}\sum_{\beta=\pm}\left(1+\beta\boldsymbol{\mathfrak{n}}\left(x\right)\boldsymbol{\cdot\sigma}\right)T_{+\beta}\left(\omega\right)\Theta\left(v_{x}\right)\\
-\dfrac{\mathbf{i}}{2}\sum_{\beta=\pm}\left(1-\beta\boldsymbol{\mathfrak{n}}\left(x\right)\boldsymbol{\cdot\sigma}\right)T_{-\beta}\left(\omega\right)\Theta\left(-v_{x}\right)\label{eq:gN}
\end{multline}
\begin{equation}
T_{\alpha\beta}\left(\omega\right)=\tan\left(\frac{\omega L}{\left|v_{x}\right|}+\arcsin\dfrac{\omega}{\Delta}+\alpha\dfrac{\varphi}{2}+\beta\dfrac{\Phi}{2}\right)\label{eq:T}
\end{equation}

The function $T_{\alpha\beta}(\omega)$ in \eqref{eq:gN} is a constant
in space, and represents the spectrum of the electronic states above
and below the energy gap. Above the gap, one has to analytically continue
the function $\arcsin\left(\omega/\Delta\right)$ such that $g\rightarrow1$
when $\omega\rightarrow\infty$. This spectral function contains the
phase shift $\Phi$ due to the spin precession when the electron and
hole propagate along an Andreev loop in the normal region. In addition,
the complete spin structure of the Green function appears as the Pauli
matrix $\boldsymbol{\mathfrak{n}\cdot\sigma}$ which can be position
dependent, as we will explore in section \ref{sec:Spin-effects}.

The electron Green's function in the superconductors reads
\begin{multline}
g\left(x\geq x_{R}\right)=\dfrac{-\mathbf{i}\omega}{\sqrt{\Delta^{2}-\omega^{2}}}\left(1-e^{-2\frac{x-x_{R}}{|v_{x}|}\sqrt{\Delta^{2}-\omega^{2}}}\right)\\
+e^{-2\frac{x-x_{R}}{|v_{x}|}\sqrt{\Delta^{2}-\omega^{2}}}g\left(x_{R}\right)
\end{multline}
\begin{multline}
g\left(x\leq x_{L}\right)=\dfrac{-\mathbf{i}\omega}{\sqrt{\Delta^{2}-\omega^{2}}}\left(1-e^{2\frac{x-x_{L}}{|v_{x}|}\sqrt{\Delta^{2}-\omega^{2}}}\right)\\
+e^{2\frac{x-x_{L}}{|v_{x}|}\sqrt{\Delta^{2}-\omega^{2}}}g\left(x_{L}\right)\label{eq:gS}
\end{multline}
with $g\left(x_{L,R}\right)$ given by Eq.\eqref{eq:gN}. As expected
physically the spin dependent component of the solution decays exponentially
from the S/N interfaces with characteristic length $\hbar v_{F}/\sqrt{\Delta^{2}-\omega^{2}}$
at a given energy, so that $g(x)$ converges to the bulk solution
$g_{\infty}=-\mathbf{i}\omega/\sqrt{\Delta^{2}-\omega^{2}}$ when
$x\rightarrow\pm\infty$. 

Given the electronic Green's function we can compute the observables
\eqref{eq:current-clean}-\eqref{eq:density-spin}. In the normal
region one obtains the following final results
\begin{equation}
j_{x}=-e\dfrac{\pi}{2}N_{0}k_{B}T\sum_{n=0}^{\infty}\sum_{\alpha,\beta=\pm}\left\langle \left|v_{x}\right|\alpha T_{\alpha\beta}\left(\mathbf{i}\omega_{n}\right)\right\rangle \label{eq:j}
\end{equation}
\begin{multline}
S^{a}\left(x\right)=-\dfrac{\pi}{2}N_{0}k_{B}T\\
\times\sum_{n=0}^{\infty}\sum_{\alpha,\beta=\pm}\left\langle \sgn\left(v_{x}\right)\mathfrak{n}^{a}\left(x\right)\beta T_{\alpha\beta}\left(\mathbf{i}\omega_{n}\right)\right\rangle \label{eq:rho-spin}
\end{multline}
\begin{multline}
\mathfrak{J}_{i}^{a}\left(x\right)=-\dfrac{\pi}{2}N_{0}k_{B}T\\
\times\sum_{n=0}^{\infty}\sum_{\alpha,\beta=\pm}\left\langle v_{i}\sgn\left(v_{x}\right)\mathfrak{n}^{a}\left(x\right)\beta T_{\alpha\beta}\left(\mathbf{i}\omega_{n}\right)\right\rangle \label{eq:j-spin}
\end{multline}
where $T_{\alpha\beta}\left(\omega\right)$ in Eq.\eqref{eq:T} is
analytically continued to imaginary axis and evaluated at the Matsubara
frequencies.

The expressions \eqref{eq:j}-\eqref{eq:j-spin} represent the central
result of the present paper. They are valid for any length, temperature
and spin dependent interaction encoded in the definitions of $\Phi$
and $\boldsymbol{\mathfrak{n}}$ in \eqref{eq:WA-expansion}. They
are also independent of the explicit form of the Fermi surface, which
is hidden in the angular averaging $\left\langle \cdots\right\rangle $.
In a general case these expressions can only be evaluated numerically.
Notice that for a ballistic S/N/S junction, when there is no magnetic
interaction in N one simply has $u=\bar{u}=1$ and thus $\Phi=0$,
$\boldsymbol{\mathfrak{n}}=\boldsymbol{0}$ and Eq. \eqref{eq:j}
gives the well known current-phase relation 
\begin{multline}
j_{x}^{\text{SNS}}=-eN_{0}k_{B}T\sum_{n=0}^{\infty}\sum_{\alpha=\pm}\times\\
\left\langle \left|v_{x}\right|\alpha\tan\left(\dfrac{\mathbf{i}\omega_{n}L}{\left|v_{x}\right|}+\arcsin\dfrac{\mathbf{i}\omega_{n}}{\Delta}+\alpha\dfrac{\varphi}{2}\right)\right\rangle \label{eq:j-SNS}
\end{multline}
which has been studied intensively in the literature, see \textit{e.g.}
\citep{nazarov.1994,nazarov.1999} and references therein. A finite
spin-orbit coupling in the S/N/S does not change the current-phase
relation since it will lead to the symmetry $u=\bar{u}$ and hence
$W=1$, which implies $\Phi=0$, see \eqref{eq:WA-expansion}. 

From the equation of motion \eqref{eq:u} for the $u$'s and the definition
\eqref{eq:WA} for $W$ one has
\begin{equation}
\mathbf{i}\dfrac{dW}{ds}+\left[A_{0}+\boldsymbol{v\cdot A},W\right]=0\label{eq:W-eq}
\end{equation}
\begin{equation}
\dfrac{d\boldsymbol{\mathfrak{n}}^{a}}{ds}+2\varepsilon^{abc}\left(A_{0}^{b}+v_{i}A_{i}^{b}\right)\boldsymbol{\mathfrak{n}}^{c}=0\label{eq:n-eq}
\end{equation}
where we have used \eqref{eq:WA-expansion} for the second expression.
One sees from these equations that the spin quantities \eqref{eq:rho-spin}
and \eqref{eq:j-spin} are covariantly conserved $\mathfrak{D}_{0}S+\mathfrak{D}_{i}\mathfrak{J}_{i}=0$
because $\left[A_{0}+\boldsymbol{v\cdot A},\boldsymbol{\mathfrak{n}\cdot\sigma}\right]^{a}=2\mathbf{i}\varepsilon^{abc}\left(A_{0}^{b}+v_{i}A_{i}^{b}\right)\boldsymbol{\mathfrak{n}}^{c}\sigma^{c}$.

According to Eq. \eqref{eq:j}, the charge current is only odd in
phase (it is odd in the sum over $\alpha$), a generic statement from
a problem when the $\text{SU}\left(2\right)$ magnetic field is neglected,
as discussed in details in \citep{Konschelle2015}. In contrary, the
spin-observables \eqref{eq:rho-spin} and \eqref{eq:j-spin} are even
in phase (even in the $\alpha$-summation), and so the spin observables
may be finite even if the supercurrent is zero and $\varphi=0$. 

Simplification of equations \eqref{eq:j}-\eqref{eq:j-spin} can be
readily obtained for the observables for temperatures close to the
critical temperature $T_{c}$ (equivalently when $\Delta/k_{B}T_{c}\ll1$)
in the long junction limit $L/\xi_{T}\gg1$ with the thermal length
$\xi_{T}=\hbar v_{F}/2\pi k_{B}T_{c}$
\begin{multline}
\lim_{T\rightarrow T_{c}}j_{x}=-\dfrac{N_{0}}{\pi}e\dfrac{\Delta^{2}}{k_{B}T_{c}}\sin\varphi\\
\times\left\langle e^{-Lv_{F}/\xi_{T}\left|v_{x}\right|}\left|v_{x}\right|\cos\Phi\right\rangle \label{eq:j-Tc}
\end{multline}
\begin{multline}
\lim_{T\rightarrow T_{c}}S^{a}=-\dfrac{N_{0}}{\pi}\dfrac{\Delta^{2}}{k_{B}T_{c}}\cos\varphi\\
\times\left\langle e^{-Lv_{F}/\xi_{T}\left|v_{x}\right|}\sgn\left(v_{x}\right)\mathfrak{n}^{a}\sin\Phi\right\rangle \label{eq:rho-spin-Tc}
\end{multline}
\begin{multline}
\lim_{T\rightarrow T_{c}}\mathfrak{J}_{i}^{a}=-\dfrac{N_{0}}{\pi}\dfrac{\Delta^{2}}{k_{B}T_{c}}\cos\varphi\\
\times\left\langle e^{-Lv_{F}/\xi_{T}\left|v_{x}\right|}v_{i}\sgn\left(v_{x}\right)\mathfrak{n}^{a}\sin\Phi\right\rangle \label{eq:j-spin-Tc}
\end{multline}
where $\mathfrak{n}^{a}$ and $\Phi$ contain some angular properties. 

Another compact expression for \eqref{eq:j} can be obtained in the
short-junction limit where contributions of order $\omega L/\left|v_{x}\right|\sim L/\xi_{T}\ll1$
are neglected (see Appendix \ref{app:Short-junction-limit}) 
\begin{equation}
\lim_{L/\xi_{T}\rightarrow0}\dfrac{j_{x}}{j_{0}}=\sum_{\alpha=\pm}\left\langle \dfrac{\left|v_{x}\right|}{v_{F}}K_{\alpha}\left(\varphi,\Phi\right)\right\rangle \label{eq:j-short-junction}
\end{equation}
\begin{equation}
K_{\alpha}=\sin\dfrac{\varphi+\alpha\Phi}{2}\tanh\left(\dfrac{\Delta}{2k_{B}T}\cos\dfrac{\varphi+\alpha\Phi}{2}\right)\label{eq:K-short}
\end{equation}
where $j_{0}=-e\pi v_{F}N_{0}\Delta/4$. One has to keep in mind that
the short junction limit $L\ll\xi_{T}$ has no influence on the ratio
between some magnetic length and the length of the junction, so it
is justified to keep the full dependency in $\Phi\sim L/\ell_{\text{pr}}$
(for instance the precession length $\ell_{\text{pr}}=2h/\hbar v_{F}$
for a monodomain ferromagnet with exchange field $h$ \citep{Buzdin1982}).
The current-phase expression \eqref{eq:j-short-junction} generalizes
well known expressions for the supercurrent in ballistic systems,
in all dimensions (see \textit{e.g.} \citep{Konschelle2008} and references
therein for the S/F/S case). The expressions for the spin observables
\eqref{eq:j-spin} and \eqref{eq:rho-spin} become in the limit $L\ll\xi_{T}$:
\begin{equation}
\lim_{L/\xi_{T}\rightarrow0}\dfrac{S^{a}}{S_{0}}=\sum_{\alpha=\pm}\left\langle \alpha\sgn\left(v_{x}\right)\mathfrak{n}^{a}K_{\alpha}\right\rangle \label{eq:rho-spin-short}
\end{equation}
\begin{equation}
\lim_{L/\xi_{T}\rightarrow0}\dfrac{\mathfrak{J}_{i}^{a}}{S_{0}}=\sum_{\alpha=\pm}\left\langle \alpha v_{i}\sgn\left(v_{x}\right)\mathfrak{n}^{a}K_{\alpha}\right\rangle \label{eq:j-spin-short}
\end{equation}
with $S_{0}=-\pi N_{0}\Delta/4$. 

Finally, in the long junction limit one has (see Appendix \ref{app:Short-junction-limit})
\begin{equation}
\lim_{L/\xi_{T}\rightarrow\infty}j_{x}=-e\dfrac{\pi}{2}N_{0}k_{B}T\left\langle \dfrac{\left|v_{x}\right|\cos\Phi}{\sinh\dfrac{\pi TL}{\left|v_{x}\right|}}\right\rangle \sin\varphi
\end{equation}
\begin{equation}
\lim_{L/\xi_{T}\rightarrow\infty}S^{a}=-\dfrac{\pi}{2}N_{0}k_{B}T\left\langle \dfrac{\sgn\left(v_{x}\right)\mathfrak{n}^{a}\left(x\right)\sin\Phi}{\sinh\dfrac{\pi TL}{\left|v_{x}\right|}}\right\rangle \cos\varphi
\end{equation}
\begin{equation}
\lim_{L/\xi_{T}\rightarrow\infty}\mathfrak{J}_{i}^{a}=-\dfrac{\pi}{2}N_{0}k_{B}T\left\langle \dfrac{v_{i}\sgn\left(v_{x}\right)\mathfrak{n}^{a}\left(x\right)\sin\Phi}{\sinh\dfrac{\pi TL}{\left|v_{x}\right|}}\right\rangle \cos\varphi\;.
\end{equation}

In the rest of this work we analyze the physical consequences of the
above general solution for the ballistic S/N/S Josephson junction
with generic spin-dependent fields. In particular we explore the behavior
of the spin-resolved density of states, the distribution of the spin
polarization, the spin and the charge currents for various configurations
of the spin fields.

\section{Density of states \label{sec:Density-of-states}}

Let us start from the analysis of the density of states in the N bridge,
in particular its dependence on the magnetic phase shift $\Phi$. 

\begin{figure}[b]
\includegraphics[width=0.49\columnwidth]{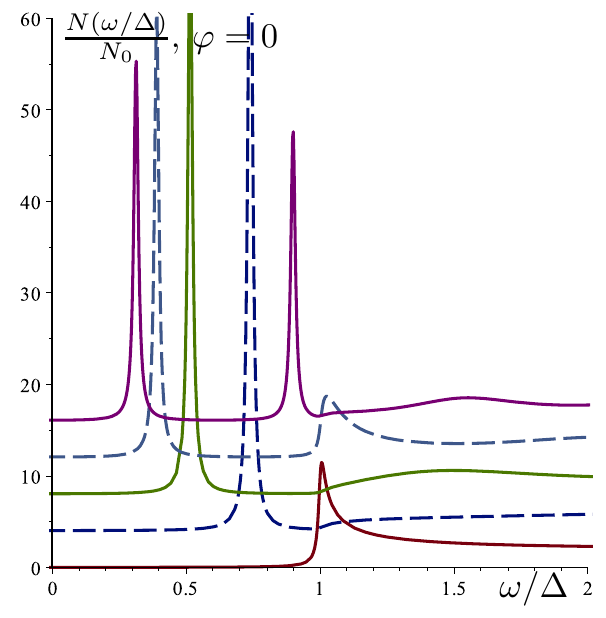}\includegraphics[width=0.49\columnwidth]{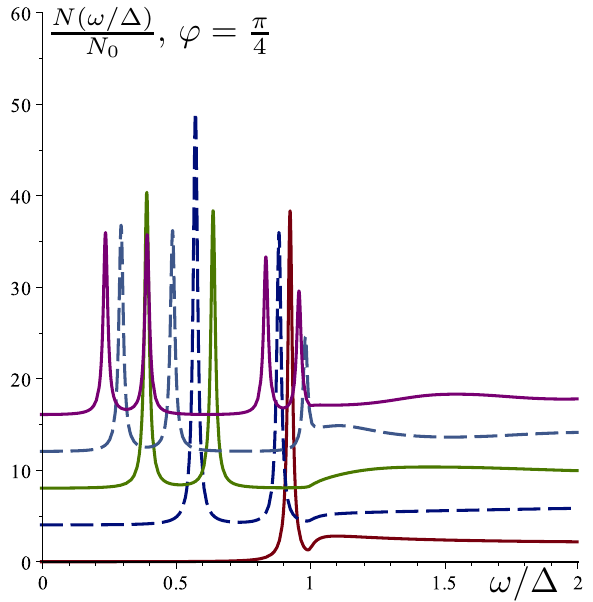}

\includegraphics[width=0.49\columnwidth]{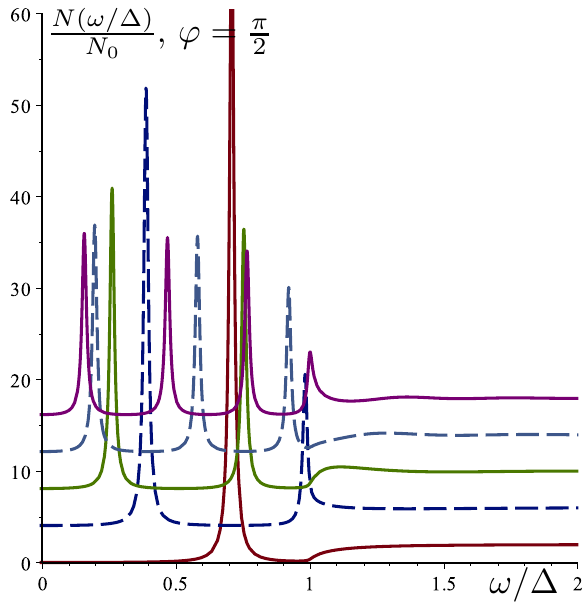}\includegraphics[width=0.49\columnwidth]{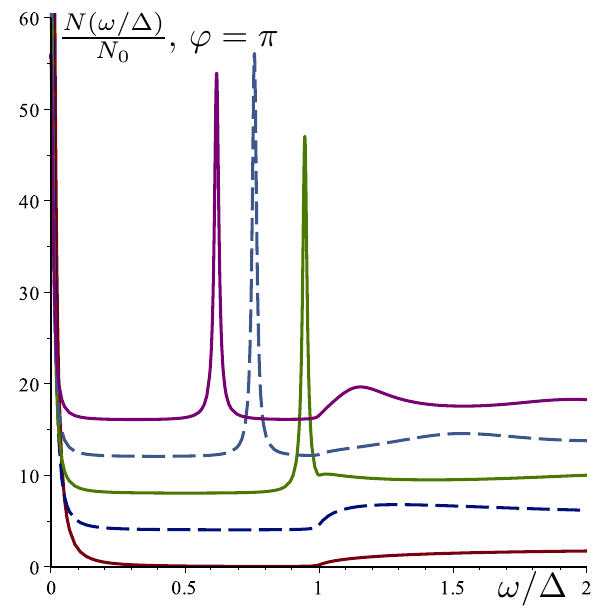}

\caption{\label{fig:Density-of-state}Density of state with respect to the
energy $\omega/\Delta$ from the expression \eqref{eq:DOS} for $\Phi=0$.
In each panel different lengths of the junction are represented: $L/\xi_{0}=\left\{ 0,1,2,3,4\right\} $
($\xi_{0}=\hbar v_{F}/\Delta$ is the coherence length), with a constant
offset of each curve from the previous one for commodity, and an alternance
of plain and dashed curves. Top-left panel: $\varphi=0$, top-right
panel: $\varphi=\pi/4$, bottom-left panel: $\varphi=\pi/2$, bottom-right
panel: $\varphi=\pi$. One sees that increasing the length of the
junction allows the opening of new Andreev modes, which eventually
goes to zero energy when $\varphi$ approach $\pi$. Longer and longer
junctions allow for more and more Andreev modes to appear. }
\end{figure}
\begin{figure}[b]
\includegraphics[width=0.7\columnwidth]{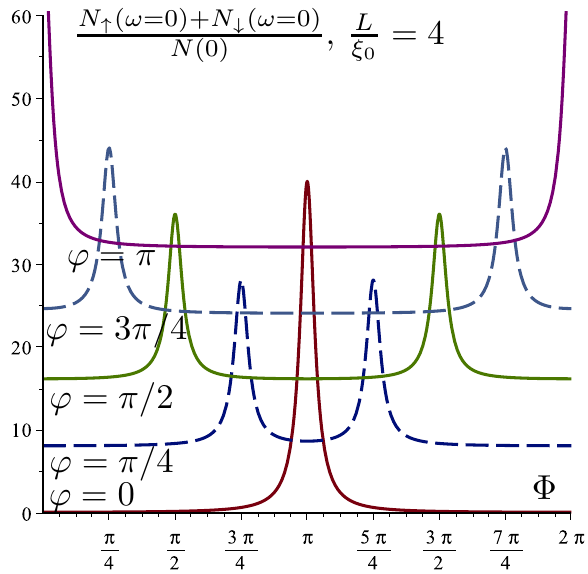}

\caption{\label{fig:zero-energy-peak} Density of state at zero energy from
the expression \eqref{eq:DOS} with respect to $\Phi$. Different
phase-difference of the junction are represented: $\varphi=\left\{ 0,1,2,3,4\right\} \times\pi/4$,
with a constant offset of each curve from the previous one for commodity,
and an alternance of plain and dashed curves. The plot is for $L/\xi_{0}=4$,
but any length displays the same zero-energy peaks, namely when $\Phi\pm\varphi=\pi$,
generally broader for longer junctions. Note that the possible length-dependency
of the parameter $\Phi$ has not been taken into account for this
check.}
\end{figure}
\begin{figure}[b]
\includegraphics[width=0.45\columnwidth]{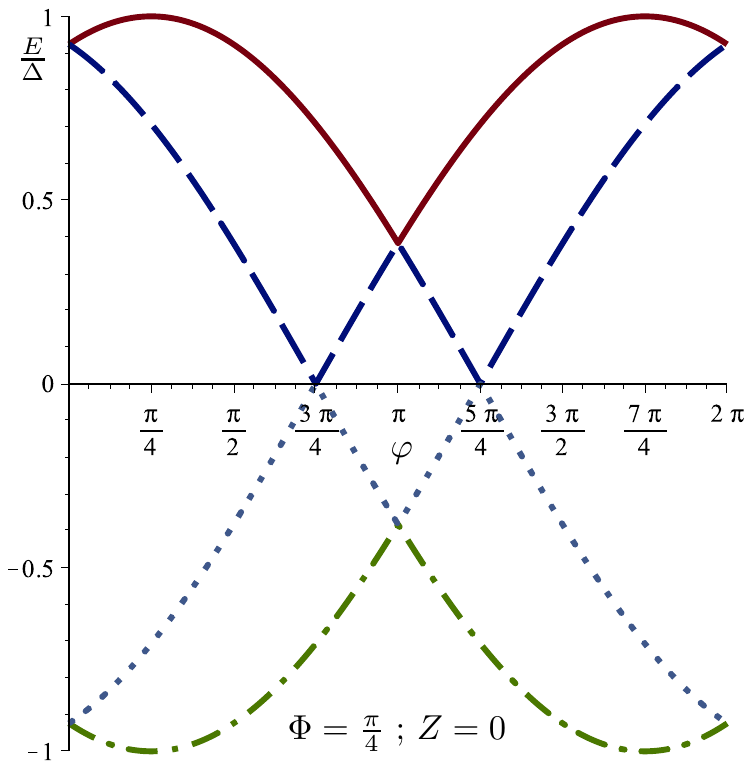}\includegraphics[width=0.45\columnwidth]{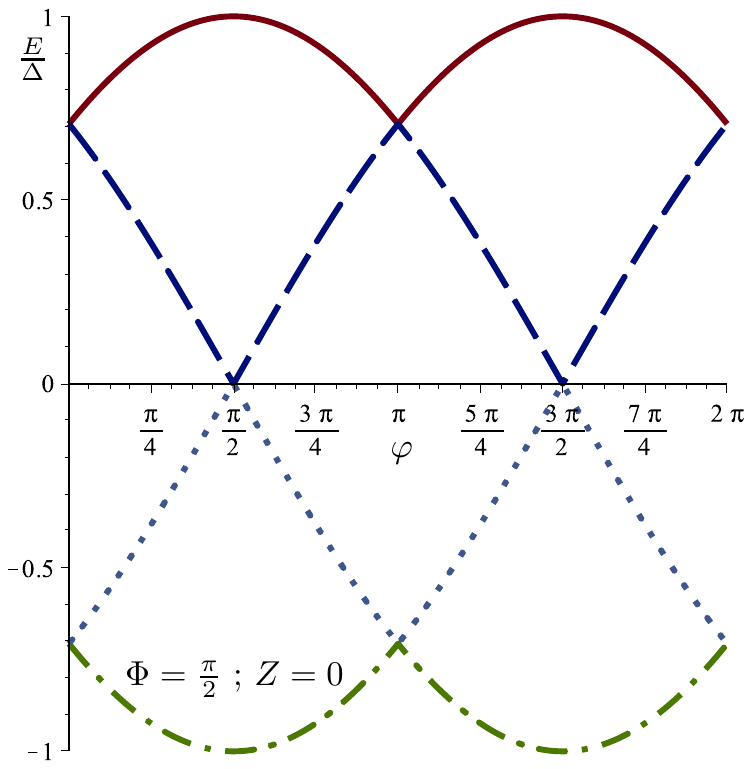}

\caption{\label{fig:Andreev} Energy $E\left(\varphi\right)/\Delta$ of the
Andreev bound states with respect to the phase difference $\varphi$
for a short junction $L/\xi_{0}\rightarrow0$, from the expression
\eqref{eq:Andreev-energy}. $\Phi=\pi/4$ (left panel), and $\Phi=\pi/2$
(right panel). At the value $\varphi=\pi$ and $\varphi=\pi\pm\Phi$,
the curves are spin degenerate in the upper or lower half planes.}
\end{figure}

From the expression for the Green's function \eqref{eq:gN}, one sees
that the local matrix structure $\boldsymbol{\mathfrak{n}\cdot\sigma}$
can be diagonalized at any given point $x$. The quantities
\begin{equation}
g_{\uparrow}(\omega)=-\dfrac{\mathbf{i}}{2}\left(T_{++}\left(\omega\right)\Theta\left(v_{x}\right)+T_{-+}\left(\omega\right)\Theta\left(-v_{x}\right)\right)
\end{equation}
and 
\begin{equation}
g_{\downarrow}(\omega)=-\dfrac{\mathbf{i}}{2}\left(T_{+-}\left(\omega\right)\Theta\left(v_{x}\right)+T_{--}\left(\omega\right)\Theta\left(-v_{x}\right)\right)
\end{equation}
represent Green's functions for the spin up and down with respect
to the local axis $\boldsymbol{\mathfrak{n}}(x)$ , with the function
$T_{\alpha\beta}(\text{\ensuremath{\omega})}$ defined in Eq. \eqref{eq:T}.
Then the density of states (DOS) per unit energy and per unit spin
is calculated as follows
\begin{equation}
\dfrac{N_{\uparrow,\downarrow}\left(\omega\right)}{N\left(0\right)}=\lim_{\epsilon\rightarrow0}\dfrac{\Re\left\langle g_{\uparrow,\downarrow}\left(\omega+\mathbf{i}\epsilon\right)\right\rangle }{\pi}\label{eq:DOS}
\end{equation}
The total DOS is given by $N_{\uparrow}+N_{\downarrow}$ whereas the
spectral spin density will be proportional to $N_{\uparrow}-N_{\downarrow}$.
These two quantities can be in principle measured by spin-polarized
near-field spectroscopy. 

The DOS is plotted on Fig.\ref{fig:Density-of-state} for different
ratios of $L/\xi_{0}$ with $\xi_{0}=\hbar v_{F}/\Delta$ the coherence
length, and for several values of $\varphi$ at $\Phi=0$. The increase
of the junction length leads to an increase of the number of Andreev
channels. A zero energy peaks in the DOS appear for $\varphi\approx\pi$.

When $\Phi\neq0$, a peak at zero-energy can be generated each time
$\Phi\pm\varphi=\pi$. For instance when $\varphi=\pi/4$, a zero-energy
peak appears for $\Phi=3\pi/4$ and for $\Phi=5\pi/4$. The spectral
weight of the Andreev channels depend on the precise value of the
total phase ($\varphi\pm\Phi$ plus the length contribution $\omega L/v_{F}$)
accumulated along the junction, and the zero-energy peaks tend to
decrease and become broader for longer junction, see Fig.\ref{fig:zero-energy-peak}.
Importantly, the zero-energy peak seems to be a generic feature of
the presence of magnetic interaction in the ballistic bridge. Zero
bias peaks were also obtained in diffusive Josephson systems \citep{Jacobsen2015,Arjoranta2015}
or in S/F/F systems \citep{Alidoust2015b}. 

The Andreev bound-states are now routinely measured
in state-of-the-art experiments \citep{Lee2012a,Bretheau2013a}.
They are obtained from the poles of the denominator in Eq.\eqref{eq:DOS}.
Specifically the energies $E_{\alpha,\beta}$ with $\alpha,\beta=\left\{ \pm\right\} $
of the bound states verify the following quasi-classic quantization
condition
\begin{equation}
\dfrac{E_{\alpha,\beta}L}{v_{x}}-\arccos\dfrac{E_{\alpha,\beta}}{\Delta}+\alpha\dfrac{\varphi}{2}+\beta\dfrac{\Phi}{2}=n\pi\;;\;n\in\mathbb{Z}\;.\label{eq:quantization-condition}
\end{equation}
The two bound states characterized by the energy $E_{\alpha+}$ and
$E_{\alpha-}$ correspond to the spin degree of freedom, and so a
spin-resolved spectroscopy of the Andreev bound states with respect
to the phase at a given length and temperature ($\Delta$ fixed) would
determine $\Phi$. The bound states are double degenerate when $\Phi=2n\pi$,
$\forall n\in\mathbb{Z}$ ; in these cases the quantization condition
\eqref{eq:quantization-condition} has been first established by Kulik
in the pure semi-classic limit, i.e. when $E/\Delta\rightarrow0$
and so $\arccos\left(E/\Delta\right)\rightarrow\pi/2$, the Maslov
index being $1/2$ in this case \citep{kulik.1970,Duncan2002}.

When $L/\xi_{0}=0$, there are four branches with expressions
\begin{equation}
\dfrac{E_{\sigma}\left(\varphi\right)}{\Delta}=\pm\sqrt{\cos^{2}\dfrac{\varphi+\sigma\Phi}{2}}\label{eq:Andreev-energy}
\end{equation}
as plotted in Fig.\ref{fig:Andreev}. They are at least double degenerate
at the value $E_{\alpha\pm}=0$ when $\varphi=\pm\Phi$. When $\Phi=0$
or $\Phi=\pi$, the two curves in the upper half-plane are spin degenerate,
as well as the two curves in the lower half-plane. Similar result
has been obtained in \citep{Fogelstrom2000,Barash2002}, eventually
generalized to the case of a point-contact with spin-active interfaces.
One sees that the zero-energy DOS obtained in Fig.\ref{fig:zero-energy-peak}
are in fact associated to the anti-crossing of the Andreev bound states
(at least in the short-junction limit). 

We should emphasize that the zero-energy Andreev
states obtained here are a generic consequence of proximity effect
under spin interactions and that they also exist for finite transparency,
as shown in our previous work \citep{Konschelle2016}. These zero
energy states are not related to zero-energy Majorana modes, which
are absent of our analysis. Our zero-energy crossings do not describe
any topological phase transition since they occur only at certain
values of the spin-splitting field encoded in $\Phi$.

\section{Magnetic Moment in a S/F/S Junction\label{sec:SFS}}

In this section we calculate the spin polarization of conduction electrons
in a S/F/S structure in the simplest non-trivial situation with $\Phi\neq0$
and a constant vector $\boldsymbol{\mathfrak{n}}$. This is, nevertheless,
a generalization of the results for the magnetic moment induced in
S/F bilayers studied in earlier works \citep{Bergeret2004,Krivoruchko2002,Bergeret2004a,Bergeret2004b,Bergeret2005,Bergeret2005a,Kharitonov2006,Xia2009a}.
The case of a Bloch domain wall, which implies a position dependent
$\boldsymbol{\mathfrak{n}}$, will be discussed in section \ref{sec:Bloch-domain-wall}. 

For definiteness, we choose $\boldsymbol{\mathfrak{n}}\left(x\right)=\boldsymbol{\hat{z}}$,
the unit vector along the $z$-axis. This situation corresponds to
a variable exchange-field directed along the $z$-axis only $A_{0}=h\left(x\right)\sigma_{z}/\hbar$,
but with arbitrary spatial dependence. In this case $\Phi=2\int_{-L/2}^{L/2}h\left(x\right)dx/v_{x}$
is the exchange field integrated over the junction. When $h$ is constant
along the junction, then $\Phi=2hL/v_{x}$ and we recover the usual
oscillations of the critical current with respect to the length and/or
exchange field of the junction \citep{buzdin.2005_RMP,Bergeret2005,Konschelle2008}.
For an anti-ferromagnetic ordering, say for example two equal domains
with up and down magnetization, $\Phi=0$ and there is no signature
of the magnetic proximity effect, as has been obtained after a long
calculation in \citep{Blanter2003}. In contrast, our method provides
a simple and clear way to understand this issue immediately. 

\begin{figure}[b]
\includegraphics[width=0.48\columnwidth]{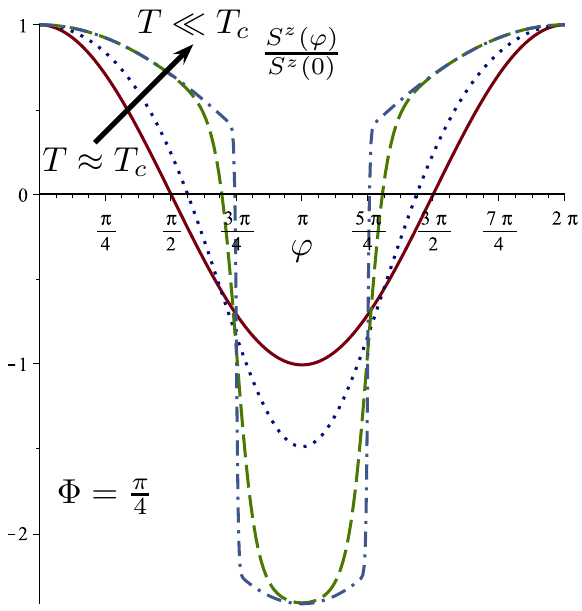}\includegraphics[width=0.48\columnwidth]{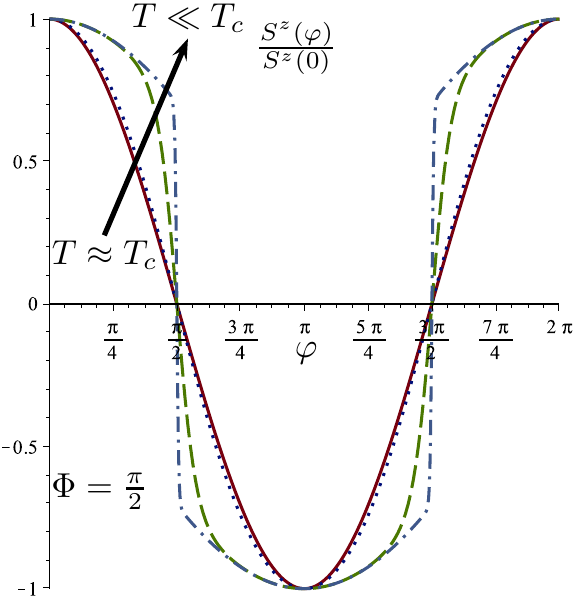}

\includegraphics[width=0.5\columnwidth]{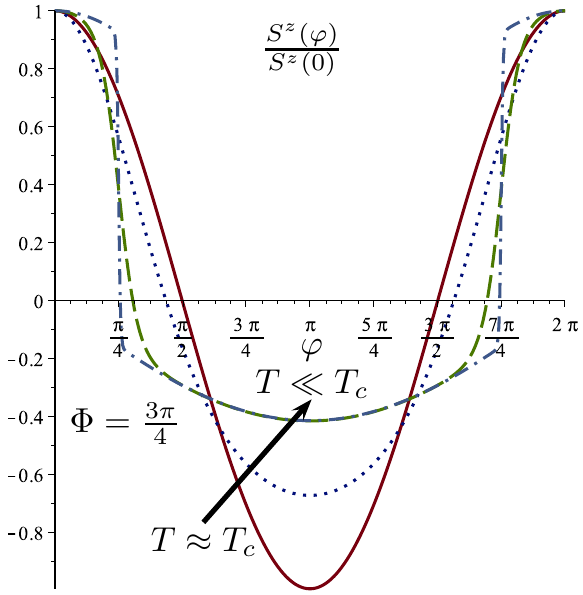}

\caption{\label{fig:spin-polar-vs-phi} Spin-polarization $S^{z}\left(\varphi\right)/S^{z}\left(0\right)$
in 1D monodomain S/F/S Josephson-junction from Eq.\eqref{eq:rho-spin-monodomain-short}
with respect to the phase-difference $\varphi$ and for different
values of the temperature $t=T/T_{c}=0.99$ (plain/red), $0.5$ (blue/dotted),
$0.1$ (green/dashed) and $0.01$ (blue/dotted-dashed) showing steeper
and steeper curves as the temperature is decreased. Left panel: $\Phi=2hL/v_{F}=\pi/4$.
Middle panel: $\Phi=\pi/2$. Right panel: $\Phi=3\pi/4$.}
\end{figure}
\begin{figure}[b]
\includegraphics[width=0.48\columnwidth]{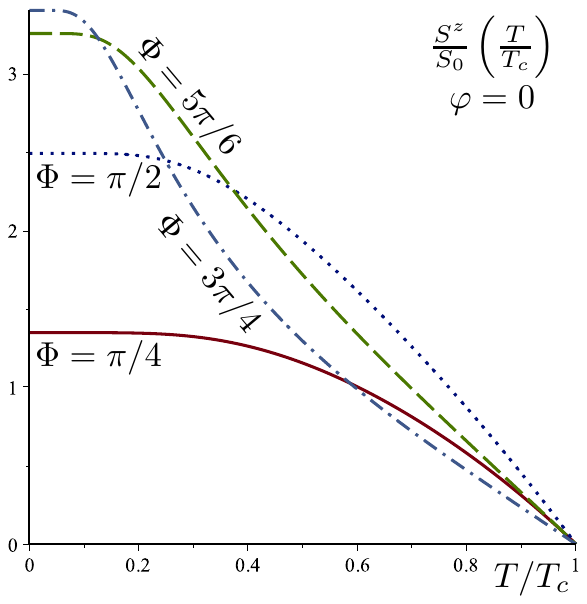}\includegraphics[width=0.48\columnwidth]{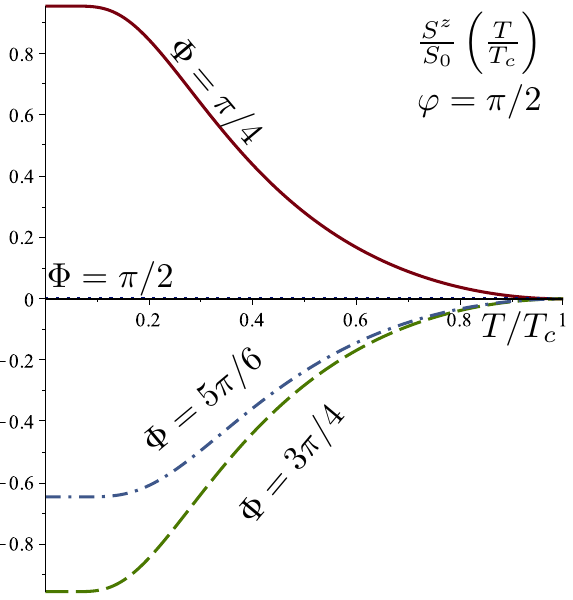}

\includegraphics[width=0.48\columnwidth]{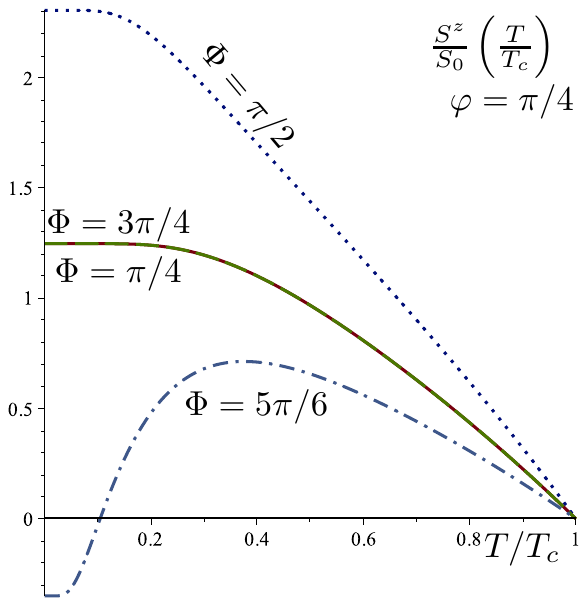}\includegraphics[width=0.48\columnwidth]{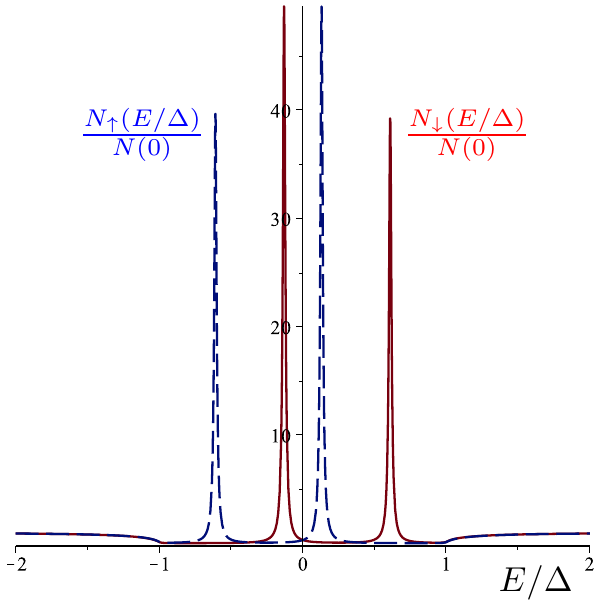}

\caption{\label{fig:spin-polar-vs-temp} Spin-polarization $S^{z}$ in 1D monodomain
S/F/S Josephson-junction from Eq.\eqref{eq:rho-spin-monodomain-short}
with respect to the temperature $T/T_{c}$ for different values of
$\Phi=2hL/v_{F}=\pi/4$ (red/plain), $\pi/2$ (blue/dotted), $3\pi/4$
(green/dashed), $5\pi/6$ (blue/dotted-dashed), at zero-phase-difference
$\varphi=0$ (upper left panel), $\pi/2$ (upper right panel) and
$\pi/4$ (lower left panel). The diamagnetic magnetization saturates
at low temperatures and vanishes for $T\rightarrow T_{c}$. A reversal
of the spin polarization is predicted in the bottom panel. This is
understood when plotting the density of state for $\varphi=\pi/4$
and $\Phi=5\pi/6$ (lower right panel), when one superimposes a Fermi
distribution law, since the heights of the different peaks are similar.}
\end{figure}

The charge current \eqref{eq:j} in the junction reduces in this case
to the form 
\begin{multline}
j_{x}^{\text{SFS}}=-e\dfrac{\pi}{2}N_{0}k_{B}T\sum_{n=0}^{\infty}\sum_{\alpha,\beta=\pm}\times\\
\left\langle \left|v_{x}\right|\alpha\tan\left(\dfrac{\mathbf{i}\omega_{n}L}{\left|v_{x}\right|}+\arcsin\dfrac{\mathbf{i}\omega_{n}}{\Delta}+\alpha\dfrac{\varphi}{2}+\beta\dfrac{\Phi}{2}\right)\right\rangle \label{eq:j-SFS}
\end{multline}
Such a current-phase relation has been thoroughly investigated in
the past, see e.g. \citep{buzdin.2005_RMP,Bergeret2005,golubov_kupriyanov.2004,Konschelle2008}

In addition to the charge current, an S/F/S junction is expected to
host a finite spin polarization, which can be calculated from Eq.\eqref{eq:rho-spin}.
When $A_{0}=h\left(x\right)\sigma_{z}/\hbar$, only $S^{z}$ survives,
and we get
\begin{equation}
S^{z}=-\dfrac{\pi}{2}N_{0}k_{B}T\sum_{n=0}^{\infty}\sum_{\alpha,\beta=\pm}\left\langle \sgn\left(v_{x}\right)\beta T_{\alpha\beta}\left(\mathbf{i}\omega_{n}\right)\right\rangle 
\end{equation}
for arbitrary length or
\begin{equation}
\lim_{L\ll\xi_{T}}\dfrac{S^{z}}{S_{0}}=\sum_{\alpha=\pm}\left\langle \alpha\sgn\left(v_{x}\right)K_{\alpha}\right\rangle \label{eq:rho-spin-monodomain-short}
\end{equation}
in the short-junction limit, with $T_{\alpha\beta}$ given in Eq.
\eqref{eq:T} and $K_{\alpha}$ in \eqref{eq:K-short}. Since we are
considering a constant $\boldsymbol{\mathfrak{n}}$, the spin-polarization
is position independent inside the bridge and vanishes for $\Phi=\left\{ 0,\pi\right\} $. 

We show in Fig.\ref{fig:spin-polar-vs-phi} the spin-polarization
in a short 1D junction ($v_{x}=\pm v_{F}$) versus the phase difference
$\varphi$ and in Fig.\ref{fig:spin-polar-vs-temp} as a function
of the temperature $T/T_{c}$. For the computation we use the interpolation
formula $\Delta\left(T\right)\approx1.764T_{c}\tanh\left(1.74\sqrt{T_{c}/T-1}\right)$
for the superconducting gap. 

It is clear from Fig.\ref{fig:spin-polar-vs-temp} that the induced
correction to the spin-polarization is a consequence of the proximity
effect and vanishes at temperatures larger than the critical temperature.
For certain values of $\varphi$ and $\Phi$ the spin-polarization
can change its sign as a function of temperature, as shown, for example,
in the right panel of Fig.\ref{fig:spin-polar-vs-temp} for $\Phi=5\pi/6$
and $\varphi=\pi/4$. This behavior can be explained by the thermal
occupancy of the Andreev levels induced in the F region. In Fig.\ref{fig:spin-polar-vs-temp}
we show the spin polarized density of state (DOS) for $\Phi=5\pi/6$
and $\varphi=\pi/4$. One sees that the peaks in the DOS for up and
down electrons have different height. The spin-polarization is obtained
by multiplying the DOS by the occupancy of the levels, i.e the equilibrium
Fermi distribution function, and integrating over energies. At low
temperatures the dominant peak is the left-most one, which is $\uparrow$-spin
polarized, resulting in a negative spin polarization. However for
higher enough temperatures, the higher peak being $\downarrow$-polarized
starts to become more populated, and at $T\rightarrow T_{c}$ its
height dominates over the $\uparrow$-peak, i.e. the surface under
the Fermi distribution favors the $\downarrow$-polarization, then
making an overall spin polarization of positive sign.

\section{Spin polarization and spin currents in junctions with spin-orbit
coupling\label{sec:Spin-effects}}

In this section we discuss the spin dependent effects in a junction
with both spin-splitting and spin-orbit effects. 
We calculate the spin density and the spin current in the normal region
of the junction. Experimental interests in these quantities grew up
recently \citep{Wakamura2014,Wakamura2015}.

To remain on analytically accessible ground, we consider the situations
when both $A_{0}$ and $\boldsymbol{v\cdot A}$ are position independent.
In that case, we can integrate \eqref{eq:u} exactly. Firstly, due
to the translational invariance in the N-region the spin propagators
depend only on the difference of coordinates: $u\left(s_{2},s_{1}\right)=u\left(s_{1}-s_{2}\right)$
and the spin propagator reduces to a simple exponential,

\[
u\left(s\right)=e^{\mathbf{i}\left(A_{0}+\boldsymbol{v\cdot A}\right)s}\equiv\exp\left[\mathbf{i}\left(A_{0}^{a}+v^{j}A_{j}^{a}\right)\sigma^{a}s\right].
\]
Therefore the general expression of Eq.\eqref{eq:WA} for the Andreew-Wilson
loop operator simplifies as follows

\begin{multline}
W(x)=e^{\frac{\mathbf{i}}{v_{x}}\left(A_{0}+\boldsymbol{v\cdot A}\right)\left(x+L/2\right)}\\
\times e^{\frac{\mathbf{i}}{v_{x}}\left(A_{0}-\boldsymbol{v\cdot A}\right)L}e^{\frac{\mathbf{i}}{v_{x}}\left(A_{0}+\boldsymbol{v\cdot A}\right)\left(L/2-x\right)}\;.\label{eq:AW-loop}
\end{multline}
It is important to notice that $W$ is a covariant object since it
determines all observables. In particular $W$ can be written in terms
of the SU(2) electric field $F_{0k}=-\mathbf{i}\left[A_{0},A_{k}\right]$
and combinations of its covariant derivatives $\mathfrak{D}_{k}F_{0j}=-\mathbf{i}\left[A_{k},F_{0j}\right]$.
To simplify further the discussion, we focus on the $T\rightarrow T_{c}$
limits of the spin density \eqref{eq:rho-spin-Tc} and spin current
\eqref{eq:j-spin-Tc}. We also restrict the analysis to the 1D and
2D situations, which are relevant experimentally. We thus consider
only Rashba ($A_{x}^{y}=-A_{y}^{x}$) or Dresselhaus ($A_{x}^{x}=-A_{y}^{y}$)
cases. According to \eqref{eq:rho-spin-Tc} and \eqref{eq:j-spin-Tc}
the spin density and spin current are determined by $\boldsymbol{\mathfrak{n}}\left(x\right)\sin\Phi$
which is the spin non-trivial part of $W$. By performing an expansion
of \eqref{eq:AW-loop} with respect to $LA_{i}/v_{x}$ one can show
that the spin density reads:\begin{widetext}
\begin{equation}
S=-\dfrac{N_{0}}{\pi}\dfrac{\Delta^{2}}{k_{B}T_{c}}\cos\varphi\left\langle e^{-Lv_{F}/\xi_{T}\left|v_{x}\right|}\sgn\left(v_{x}\right)\left[\frac{2L}{v_{x}}A_{0}+\frac{2xL}{v_{x}^{2}}v_{k}F_{0k}-\frac{4L^{3}}{3v_{x}^{3}}A_{0}^{3}-\dfrac{L}{v_{x}^{3}}v_{j}\mathfrak{D}_{j}F_{0k}v_{k}\left(\frac{L^{2}}{12}+x^{2}\right)+\cdots\right]\right\rangle \label{eq:S-expans}
\end{equation}
whereas the spin current takes the form 
\begin{equation}
\mathfrak{J}_{i}=-\dfrac{N_{0}}{\pi}\dfrac{\Delta^{2}}{k_{B}T_{c}}\cos\varphi\left\langle e^{-Lv_{F}/\xi_{T}\left|v_{x}\right|}v_{i}\sgn\left(v_{x}\right)\left[\left(\dfrac{L}{v_{x}^{3}}\mathfrak{D}_{0}F_{0k}v_{k}-\dfrac{2}{3}\dfrac{xL}{v_{x}^{4}}\mathfrak{D}_{0}v_{j}\mathfrak{D}_{j}F_{0k}v_{k}\right)\left(\frac{L^{2}}{4}-x^{2}\right)+\cdots\right]\right\rangle \label{eq:J-expans}
\end{equation}

\end{widetext}

We now discuss the physical meaning of \eqref{eq:S-expans} and \eqref{eq:J-expans}.
For a short enough junction, the surviving term is the $S\propto A_{0}L$,
which is nothing but the polarization caused by the spin-splitting
field in S/F/S system, see section \ref{sec:SFS}. 

\begin{figure}[b]
\includegraphics[width=0.65\columnwidth]{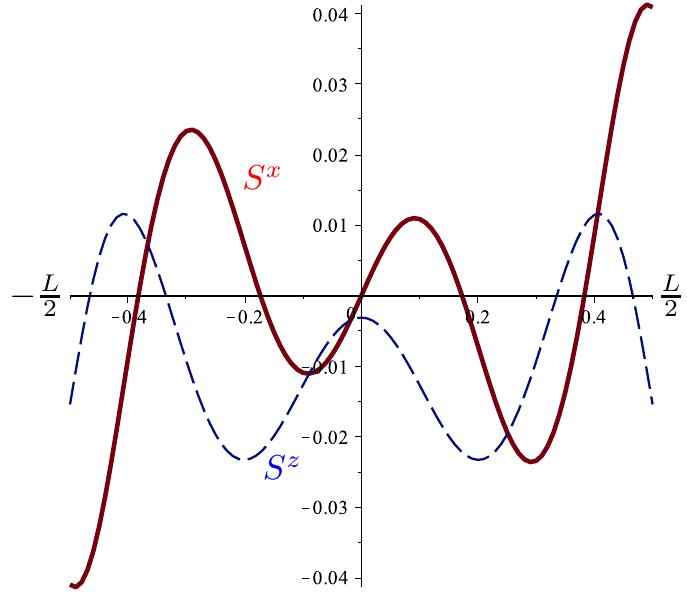}

\includegraphics[width=0.65\columnwidth]{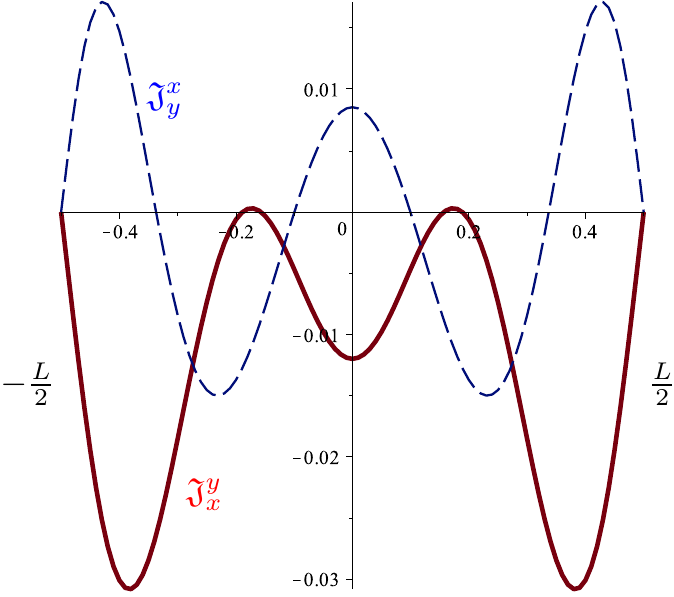}

\caption{\label{fig:spin-capacitor} The upper panel shows the spatial dependence
of the spin polarizations $S^{x}$ (plain curve) and $S^{z}$ (dashed
curve) for a Rashba SOC and an exchange field in $z$-direction. Lower
panel shows the spatial dependence of the spin currents $\mathfrak{J}_{x}^{y}$
(plain curve) and $\mathfrak{J}_{y}^{x}$ (dotted curve) in the same
situation. with $\mathfrak{J}_{i}\propto\mathfrak{D}_{0}F_{0i}$ in
analogy with a displacement current, see \eqref{eq:J-expans}. The
spin capacitor effect is illustrated by the $S^{x}$ curve, being
odd in space and originating from the term $S\propto\mathfrak{D}_{i}F_{i0}$
in \eqref{eq:S-expans}. We have chosen $\boldsymbol{h}=\left(0,0,5\right)\hbar v_{F}/\xi_{T}$,
$\xi_{T}r=3$, $L/\xi_{T}=1$, and the curves are normalized by the
quantity $2N_{0}\Delta^{2}\cos\varphi/\pi k_{B}T_{c}$. For a Dresselhaus
SOC ($\gamma=\pi/2$), one obtains similar curves by making the changes
$S^{x}\rightarrow-S^{y}$, $S^{z}\rightarrow S^{z}$, $\mathfrak{J}_{x}^{y}\rightarrow\mathfrak{J}_{x}^{x}$
and $\mathfrak{J}_{y}^{x}\rightarrow\mathfrak{J}_{y}^{y}$.}
\end{figure}

In addition to the S/F/S phenomenology, there are extra phenomenologies
mixing spin-orbit and spin-splitting effects, all proportional to
the electric field $F_{0k}v_{k}$. For instance the second order term
is proportional to the electric field $xLF_{0x}$ (after angular averaging,
only $F_{0x}$ survives). In addition it is odd in space, therefore
the spin density will present different signs at the two interfaces.
Thus the term $S\propto xLF_{0x}$ can be seen as the analog to the
capacitor effect in electrostatics, here in a S/N/S junction: the
existence of a finite electric field in the N region separates ``charges''
which in this case translate into spin densities with different signs.
In other words, in the S/N/S structure the accumulation of charge
corresponds to an accumulation of the spin polarization at the boundaries
between the normal and the superconducting regions. For this reason
we denote this effect the \textit{spin capacitor effect}. It is illustrated
on Fig.\ref{fig:spin-capacitor} for a 2D Rashba system. 

The spin current has in the leading order a contribution of the type
(see Eq.\eqref{eq:J-expans}) : $\mathfrak{J}_{i}\sim\left(L^{2}/4-x^{2}\right)\mathfrak{D}_{0}F_{0i}$
(angular averaging evaluated). This resembles the expression for the
displacements currents in electrostatics as the time derivative of
the electric field, and one calls them the \textit{displacement spin
currents}. For either a Rashba or a Dresselhaus coupling, there are
potentially two components of the electric field $F_{0x}\propto\left[A_{0},A_{x}\right]$
and $F_{0y}\propto\left[A_{0},A_{y}\right]$. Each of these allow
for a displacement current along the junction or perpendicular to
it, see Fig.\ref{fig:spin-capacitor} for an illustration of $\mathfrak{J}_{x}^{y}$
and $\mathfrak{J}_{y}^{x}$ in the case of Rashba coupling, when $A_{0}^{z}$,
$A_{x}^{y}$ and $A_{y}^{x}$ are present.

\begin{figure}[b]
\includegraphics[width=0.65\columnwidth]{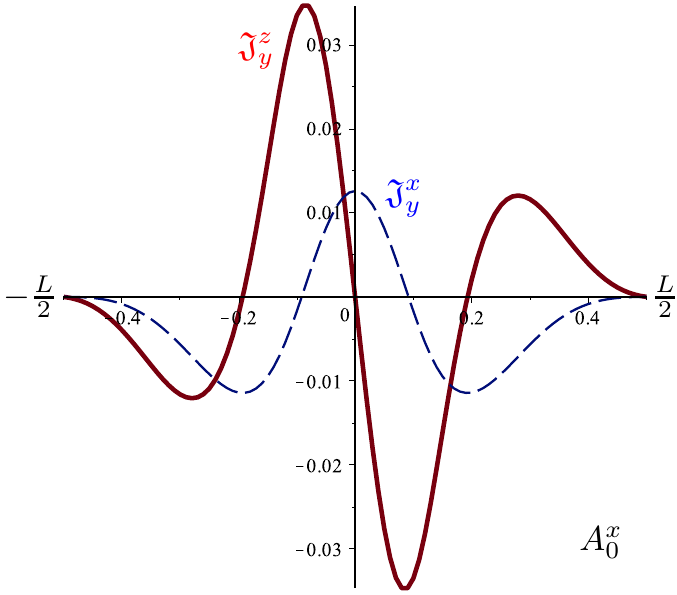}

\includegraphics[width=0.65\columnwidth]{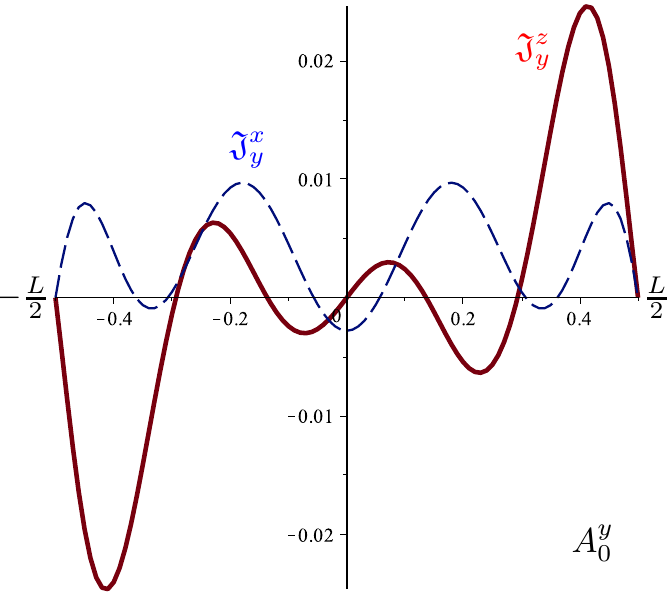}

\caption{\label{fig:transverse-current}Illustration of the transverse spin
current $\mathfrak{J}_{y}^{z,x}$ (plain/dashed curves respectively)
along the axis perpendicular to the junction for a Rashba spin-orbit
effect with respect to the junction length. We choose $L/\xi_{T}=1$,
$\xi_{T}r=3$, $\xi_{T}h_{x}=5\hbar v_{F}$ (upper panel) and $\xi_{T}h_{y}=5\hbar v_{F}$
(lower panel) in Eqs.\eqref{eq:rho-spin-Tc}-\eqref{eq:j-spin-Tc}
after injection of $\boldsymbol{\mathfrak{n}}$ from \eqref{eq:n}
and \eqref{eq:kappa}. The antisymmetric $\mathfrak{J}_{y}^{z}$ comes
from the contribution $\left\langle v_{i}\sgn\left(v_{x}\right)\mathfrak{D}_{0}v_{j}\mathfrak{D}_{j}F_{0k}v_{k}\right\rangle $
in \eqref{eq:J-expans}. The vertical axis is in unit of $2N_{0}\Delta^{2}\cos\varphi/\pi k_{B}T_{c}$.
When $h_{x}=h_{y}=0$ but $h_{z}\protect\neq0$ this term is zero,
and no asymmetric spin current is flowing, see Fig.\ref{fig:spin-capacitor}.
The current $\mathfrak{J}_{y}^{x}$ stems for higher order terms in
the expansion \eqref{eq:J-expans}.}
\end{figure}

At first sight one may think that the spin density and spin currents
may obey equations equivalent to those in electrostatics. However
this is not always true. Higher order terms in Eq. \eqref{eq:J-expans}
clearly show that the expressions for the spin density and spin current
also contains higher order covariant derivatives of the electric field
that eventually leads to creation of other components of both quantities
as discussed below. For instance the fourth order term in the expansion
Eq. \eqref{eq:J-expans} after angular averaging for a 2D gas reads
\begin{align}
\mathfrak{J}_{x}^{\left(4\right)} & \propto\mathfrak{D}_{0}\left(2\mathfrak{D}_{x}F_{x0}+\mathfrak{D}_{y}F_{y0}\right)\nonumber \\
\mathfrak{J}_{y}^{\left(4\right)} & \propto\mathfrak{D}_{0}\left(\mathfrak{D}_{y}F_{x0}+\mathfrak{D}_{x}F_{y0}\right)\;.\label{eq:Jy}
\end{align}
 In the case of Rashba SOC when the Zeeman term points toward the
$x$-axis, the only surviving term is $\mathfrak{J}_{y}^{\left(4\right)}\propto\mathfrak{D}_{0}\mathfrak{D}_{y}F_{x0}\rightarrow\mathfrak{J}_{y}^{z}$.
When the Zeeman effect is present through $A_{0}^{y}$ only, $\mathfrak{J}_{y}^{\left(4\right)}\propto\mathfrak{D}_{0}\mathfrak{D}_{x}F_{y0}$
and there are two contributions $\mathfrak{J}_{y}^{z}$ and $\mathfrak{J}_{y}^{x}$.
These situations are illustrated on Fig.\ref{fig:transverse-current}
which, as discussed below, is obtained form the exact expression.
The component of the spin current $\mathfrak{J}_{y}^{x}$ in the upper
panel of this figure would require even higher order terms in the
expansion \eqref{eq:J-expans}.

We now go beyond the above expansion and write explicitly $W$ by
combining Eq. \eqref{eq:AW-loop} with the representation Eq.\eqref{eq:WA-expansion}.
In particular we obtain the following equations which determine the
local spin quantization axis $\boldsymbol{\mathfrak{n}}\left(x\right)$
and the magnetic phase shift $\Phi$ 
\begin{multline}
\cos\Phi=\cos\dfrac{\kappa_{+}L}{v_{x}}\cos\dfrac{\kappa_{-}L}{v_{x}}\\
-\dfrac{\boldsymbol{\kappa_{+}\cdot\kappa_{-}}}{\kappa_{+}\kappa_{-}}\sin\dfrac{\kappa_{+}L}{v_{x}}\sin\dfrac{\kappa_{-}L}{v_{x}}\;,\label{eq:cosPHI}
\end{multline}
\begin{widetext}
\begin{multline}
\boldsymbol{\mathfrak{n}}\left(x\right)\sin\Phi=-\dfrac{\boldsymbol{\kappa_{+}}}{\kappa_{+}}\left(\sin\dfrac{\kappa_{+}L}{v_{x}}\cos\dfrac{\kappa_{-}L}{v_{x}}+\dfrac{\boldsymbol{\kappa_{+}\cdot\kappa_{-}}}{\kappa_{+}\kappa_{-}}\cos\dfrac{\kappa_{+}L}{v_{x}}\sin\dfrac{\kappa_{-}L}{v_{x}}\right)\\
+\left[\left(\dfrac{\boldsymbol{\kappa_{+}}}{\kappa_{+}}\dfrac{\boldsymbol{\kappa_{+}\cdot\kappa_{-}}}{\kappa_{+}\kappa_{-}}-\dfrac{\boldsymbol{\kappa_{-}}}{\kappa_{-}}\right)\cos\dfrac{2\kappa_{+}x}{v_{x}}-\dfrac{\boldsymbol{\kappa_{+}\times\kappa_{-}}}{\kappa_{+}\kappa_{-}}\sin\dfrac{2\kappa_{+}x}{v_{x}}\right]\sin\dfrac{\kappa_{-}L}{v_{x}}\label{eq:n}
\end{multline}
\end{widetext}where $\boldsymbol{\kappa_{\pm}}$ are the vectors
with the components 
\begin{equation}
\kappa_{\pm}^{a}=A_{0}^{a}\pm\sum_{i=1}^{3}v_{i}A_{i}^{a}
\end{equation}
and the norm $\kappa_{\pm}=\sqrt{\boldsymbol{\kappa_{\pm}\cdot\kappa_{\pm}}}$.
In the special case when $\boldsymbol{\kappa_{+}}=\boldsymbol{\kappa_{-}}=\boldsymbol{\kappa}$
we get $\Phi=2\kappa L/v_{x}$ and $\boldsymbol{\mathfrak{n}}\left(x\right)=\boldsymbol{\kappa}/\kappa$
(a constant), which corresponds to a pure exchange field, $A_{0}\ne0$
and $A_{i}^{a}=0$, discussed in Sec.\ref{sec:SFS}. A generic situation
will exhibit a space dependent precession axis $\boldsymbol{\mathfrak{n}}\left(x\right)$,
as can be seen from the second line of Eq.\eqref{eq:n}. In particular,
the function $\boldsymbol{\mathfrak{n}}(x)$ has both odd and even
contributions with respect to the center of the junction. Notice that
both the components of $S$ and the current can be explained for this
particular case in terms of the spin capacitor effect and displacement
currents.

We now focus on the Rashba and Dresselhaus SOC that we parametrize
using two parameters $r$ and $\gamma$. The Rashba coupling enters
as $A_{x}^{y}=-A_{y}^{x}=r\cos\gamma$ whereas the Dresselhaus coupling
reads $A_{x}^{x}=-A_{y}^{y}=r\sin\gamma$. In addition to the spin-orbit
couplings we assume a spin-splitting field parameterized by the coordinates
$h_{x,y,z}$ . One then obtains
\begin{equation}
\boldsymbol{\kappa_{\pm}}=\left(\begin{array}{c}
h_{x}\mp v_{F}r\sin\left(\phi-\gamma\right)\\
h_{y}\pm v_{F}r\cos\left(\phi+\gamma\right)\\
h_{z}
\end{array}\right)\label{eq:kappa}
\end{equation}
for a circular Fermi level parameterized by $\boldsymbol{v}=v_{F}\left(\cos\phi,\sin\phi\right)$
in a 2D system. We substitute this $\boldsymbol{\kappa}_{\pm}$ in
\eqref{eq:n} in order to evaluate the spin polarizations \eqref{eq:rho-spin-Tc}
and the spin currents \eqref{eq:j-spin-Tc} close to the critical
temperature. Fig.\ref{fig:spin-capacitor} shows the result for a
Rashba spin-orbit coupling, i.e. $\gamma=0$ in \eqref{eq:kappa}.
If one chooses a Dresselhaus coupling instead (with $\gamma=\pi/2$),
the curves are similar, except $S^{y}\left(\gamma=\pi/2\right)=-S^{x}\left(\gamma=0\right)$,
$S^{z}\left(\pi/2\right)=S^{z}\left(0\right)$, $\mathfrak{J}_{x}^{x}\left(\pi/2\right)=\mathfrak{J}_{x}^{y}\left(0\right)$,
$\mathfrak{J}_{y}^{y}\left(\pi/2\right)=\mathfrak{J}_{y}^{x}\left(0\right)$,
all the other spin observables being zero when $h_{x}=h_{y}=0$. For
this reason we do not show the case of a Dresselhaus coupling.

As expected from our previous perturbative analysis the spin density
has in both cases two contributions: the one due to the Zeeman polarization,
$S^{z}$ and those due to the capacitor effect, $S^{^{x}}$ for Rashba
and $S^{y}$ for Dresselhaus. Other contributions parallel to the
field component $F_{0y}$ vanish due to the velocity average (cf.
Eq. \ref{eq:S-expans}). Also in Fig.\ref{fig:spin-capacitor} the
currents can be explained from the lowest term in the expansion Eq.\eqref{eq:J-expans}
as displacement currents induced by electric field component $F_{0x}$,
$\mathfrak{J}_{x}^{y}$ for the Rashba case and $\mathfrak{J}_{y}^{y}$
in the Dresselhaus SOC. There are also displacements currents induced
by $F_{0y}$ component of the field which are $\mathfrak{J}_{y}^{x}$
in the Rashba case (see Fig. \ref{fig:spin-capacitor} bottom panel)
and $\mathfrak{J}_{y}^{y}$ in the Dresselhaus case (not shown). 

All the previous spin densities and currents can be explained again
in terms of the spin capacitor effect and displacement currents, and
their symmetry with respect to $x$ is determined by the leading terms
in the expansions Eqs.\eqref{eq:S-expans}-\eqref{eq:J-expans}. When
the spin-splitting field is applied in $x$ or $y$ direction also
higher order terms in Eq. \eqref{eq:J-expans} contribute to the spin
currents and generates additional components. As an example we show
the transverse currents for different directions of $A_{0}$ in Fig.
\ref{fig:transverse-current}. 

\section{Non-homogenoues exchange field \label{sec:Bloch-domain-wall}}

\begin{figure}[b]
\includegraphics[width=0.8\columnwidth]{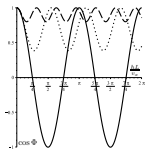}

\caption{\label{fig:cosPHI-Bloch} Magnetic interaction $\cos\Phi$ for a Bloch
domain wall with respect to the parameter $hL/\left|v_{x}\right|$,
from \eqref{eq:cosPHI-Bloch}, and for different values of $\kappa=v_{x}q/2h$.
Plain line: $\kappa=0$. Dotted line $\kappa=1.5$. Dashed line $\kappa=3$.}
\end{figure}

For completeness in this section we briefly discuss the effect of
a inhomogeneous magnetization. In particular we consider a Bloch domain
wall, characterized by a $x$-dependent exchange field
\begin{equation}
A_{0}=h\left[\sigma^{2}\sin qx+\sigma^{3}\cos qx\right]
\end{equation}
which could be included in \eqref{eq:u} in order to obtain $u\left(x\right)$.
Nevertheless, one can take advantage of the gauge-covariant formalism,
and show that the Bloch domain wall is in fact gauge-equivalent to
a situation with a constant exchange field $A_{0}^{z}=h$ in addition
to a spin-orbit interaction $A_{x}^{x}=-q/2$ \citep{Bergeret2014},
for which \eqref{eq:u} reduces to 
\begin{equation}
\mathbf{i}\dfrac{du}{ds}+\left(h\sigma^{3}-v_{x}\dfrac{q}{2}\sigma^{1}\right)u\left(s\right)=0
\end{equation}
and can be easily integrated. Since $\Phi$ is gauge-invariant, one
has 
\begin{multline}
\cos\Phi=\cos^{2}\left(\sqrt{1+\kappa^{2}}\dfrac{hL}{v_{x}}\right)-\\
-\dfrac{1-\kappa^{2}}{1+\kappa^{2}}\sin^{2}\left(\sqrt{1+\kappa^{2}}\dfrac{hL}{v_{x}}\right)\label{eq:cosPHI-Bloch}
\end{multline}
from \eqref{eq:cosPHI}, with $\kappa=v_{x}q/2h$. Expression \eqref{eq:cosPHI-Bloch}
is plotted on Fig.\ref{fig:cosPHI-Bloch}. For a monodomain, $\kappa\rightarrow0$
and $\Phi\rightarrow2hL/v_{x}$, as usual for a S/F/S junction, see
section \ref{sec:SFS}. For larger $\kappa$, $\Phi$ takes only limited
values (see Fig.\ref{fig:cosPHI-Bloch}), and for $\kappa\rightarrow\infty$
one has $\Phi\rightarrow0$, recovering a pure S/N system. This later
situation corresponds to the situation described in section \ref{sec:SFS},
when the alternance of domains with opposite spin orientations reduces
the characteristic oscillations of the S/F proximity effect, eventually
destroying these oscillations in ballistic systems when the magnetization
averaged along the junction vanishes. Since the domain wave-length
$q=-2A_{x}^{x}$ is equivalent to a spin-orbit effect, a large $\kappa$
is equivalent to a large spin-orbit effect, or equivalently a vanishing
exchange field. 

In a junction with a Bloch domain wall, there are the generation of
spin current polarized along the junction axis, as can be drawn from
the conclusions of section \ref{sec:Spin-effects} when $A_{0}^{z}$
and $A_{x}^{x}$ are present. In the lowest order in the fields, the
spin capacitor effect is present with a contribution $S^{y}\propto\partial_{x}F_{0x}^{y}$
odd in space, and the displacement spin current $\mathfrak{J}_{x}^{x}\propto\left(\mathfrak{D}_{0}F_{0x}\right)^{x}\propto A_{0}^{z}F_{0x}^{y}\left[\sigma^{z},\sigma^{y}\right]$
shows up.

\section{Conclusion}

Ballistic S/N/S Josephson systems when the normal region N exhibits
generic spin-dependent fields have been investigated. We propose a
systematic approach to study systems exhibiting both spin-splitting
and spin-orbit interactions, provided the latter is linear-in-momentum
and the magnetic interaction is weak, such that the quasi-classical
approximation is valid (section \ref{sec:Equation-of-motion}).

We have shown that the magnetic interactions appears in all observables
as a global phase accumulation $\Phi$ (see \eqref{eq:cosPhi}) and
a space dependent unit vector $\boldsymbol{\mathfrak{n}}$ (see \eqref{eq:WA-expansion}).
With the help of the derived compact expression for the quasi-classic
Green's function, Eq.\eqref{eq:gN}, we studied different spin-dependent
fields and their effects on the properties of the junction.

In particular we have demonstrated that the density of states may
show a zero-energy peak which is a generic consequence of a finite
$\Phi$.

We have also shown how such fields in the N region generate finite
changes of the spin-polarization and finite spin currents. We identify
the possibility for the accumulation of the spin at the interfaces
between the normal and superconducting regions, an effect reminiscent
to the charge accumulation at the plates of a capacitor. Hence we
call this phenomenon the spin capacitor effect. In addition, we predict
the generation of spin currents flowing along the superconducting
interfaces. Both these effects can be understood in a convenient way
using an $\text{SU}\left(2\right)$ electrostatics, which generalizes
the Maxwell electrostatics to the non-Abelian case.

Effects like the spin capacitor, or the predicted spectral features
 can be experimentally verified in superconducting heterostructures
which are being fabricated in the present, and attract more and more
interest recently. The measurement of the spin polarization and charge
current can serve as a powerful characterization of the symmetries
of the spin texture. The tunability of the superconducting condensate
via voltage or current bias allow for a coherent manipulation of the
spin polarization and currents. Reciprocally, the manipulation of
the spin quantities allow for the manipulation of the superconducting
coherent states. Research along these lines are promising in addition
to the search for topological effects in spin textured superconducting
systems.
\begin{acknowledgments}
We thank D. Bercioux and V.N. Golovach for remarks. F.K. thanks F.
Hassler and G. Viola for daily stimulating discussions during his
time at the IQI-RWTH Aachen. Special thanks are also due to A.I. Buzdin
and A. Larat. 

F.K. is grateful for support from the Alexander von Humboldt foundation.
The work of F.K. and F.S.B. was supported by Spanish Ministerio de
Economia y Competitividad (MINECO) through the Project No. FIS2014-55987-P
and the Basque Government under UPV/EHU Project No. IT-756-13. I.V.T.
acknowledges support from the Spanish Grant FIS2013-46159-C3-1-P,
and from the \textquotedblleft Grupos Consolidados UPV/EHU del Gobierno
Vasco\textquotedblright{} (Grant No. IT578-13)
\end{acknowledgments}

\appendix

\section{Green's functions\label{app:Greens-functions}}

We fix positive velocities in this appendix. Let us write \eqref{eq:continuity-LR}
in its explicit form
\begin{equation}
\left(\begin{array}{cc}
Q_{11} & Q_{12}\\
Q_{21} & Q_{22}
\end{array}\right)\left(\begin{array}{cc}
-1 & g_{1}\\
0 & 1
\end{array}\right)=\left(\begin{array}{cc}
-1 & 0\\
g_{2} & 1
\end{array}\right)\left(\begin{array}{cc}
Q_{11} & Q_{12}\\
Q_{21} & Q_{22}
\end{array}\right)
\end{equation}
in the Nambu space, where the quantities $Q_{ij}$ are matrices in
the spin space, as well as the $g_{1,2}$, and can be easilly obtained
from the definition \eqref{eq:Q}. Component by component, one gets
\begin{align}
Q_{11} & =Q_{11}\nonumber \\
Q_{11}g_{1}+Q_{12} & =-Q_{12}\nonumber \\
-Q_{21} & =g_{2}Q_{11}+Q_{21}\nonumber \\
Q_{21}g_{1}+Q_{22} & =g_{2}Q_{12}+Q_{22}
\end{align}
and one sees that the solutions of the two intermediary equations
$g_{1}=-2Q_{11}^{-1}Q_{12}$ and $g_{2}=-2Q_{21}Q_{11}^{-1}$ automatically
verifies the last one, which can be thought as a consistency equation.
To get $g_{1}$ or $g_{2}$, we now want to invert $Q_{11}$, which
is a $2\times2$ matrix. Defining
\begin{equation}
\tilde{\chi}=\arcsin\dfrac{\omega}{\Delta}+\dfrac{\varphi}{2}+\omega\left(s_{R}-s_{L}\right)
\end{equation}
 one has 
\begin{equation}
2Q_{11}\cos\arcsin\dfrac{\omega}{\Delta}=e^{\mathbf{i}\tilde{\chi}}u\left(s_{R},s_{L}\right)+e^{-\mathbf{i}\tilde{\chi}}\bar{u}\left(s_{R},s_{L}\right)
\end{equation}
for the matrix $Q_{11}$ defined in \eqref{eq:Q}. The inverse $Q_{11}^{-1}$
is obtained using the property\begin{widetext} 
\begin{equation}
\left(e^{\mathbf{i}\tilde{\chi}}u\left(s_{R},s_{L}\right)+e^{-\mathbf{i}\tilde{\chi}}\bar{u}\left(s_{R},s_{L}\right)\right)\left(e^{\mathbf{i}\tilde{\chi}}u\left(s_{L},s_{R}\right)+e^{-\mathbf{i}\tilde{\chi}}\bar{u}\left(s_{L},s_{R}\right)\right)=2\cos2\tilde{\chi}+\Tr\left\{ u\left(s_{R},s_{L}\right)\bar{u}\left(s_{L},s_{R}\right)\right\} 
\end{equation}
since the $u$'s are $2\times2$ unitary matrices and thus reads $u=\dfrac{1}{\sqrt{\left|\alpha\right|^{2}+\left|\beta\right|^{2}}}\left(\begin{array}{cc}
\alpha & \beta\\
-\beta^{\ast} & \alpha^{\ast}
\end{array}\right)$ for $\alpha,\;\beta\in\mathbb{C}$. They thus verify $u+u^{\dagger}=\mathbb{I}\Tr\left\{ u\right\} $
with $\mathbb{I}$ the identity matrix. It is clear that $u\left(s_{R},s_{L}\right)\bar{u}\left(s_{L},s_{R}\right)$
is unitary as well. One has thus 
\begin{equation}
Q_{11}^{-1}=\left(e^{\mathbf{i}\tilde{\chi}}u\left(s_{L},s_{R}\right)+e^{-\mathbf{i}\tilde{\chi}}\bar{u}\left(s_{L},s_{R}\right)\right)\dfrac{2\cos\arcsin\dfrac{\omega}{\Delta}}{2\cos2\tilde{\chi}+\Tr\left\{ u\left(s_{R},s_{L}\right)\bar{u}\left(s_{L},s_{R}\right)\right\} }
\end{equation}
and finally one obtains 
\begin{equation}
g_{1}=2\mathbf{i}\dfrac{e^{-\mathbf{i}\arcsin\frac{\omega}{\Delta}}\bar{u}\left(s_{L},s_{R}\right)u\left(s_{R},s_{L}\right)-e^{\mathbf{i}\arcsin\frac{\omega}{\Delta}}u\left(s_{L},s_{R}\right)\bar{u}\left(s_{R},s_{L}\right)+2\mathbf{i}\sin\left(2\omega\left(s_{R}-s_{L}\right)+\varphi+\arcsin\dfrac{\omega}{\Delta}\right)}{2\cos\left(2\omega\left(s_{R}-s_{L}\right)+\varphi+2\arcsin\dfrac{\omega}{\Delta}\right)+\Tr\left\{ u\left(s_{R},s_{L}\right)\bar{u}\left(s_{L},s_{R}\right)\right\} }
\end{equation}
\begin{equation}
g_{2}=-2\mathbf{i}\dfrac{e^{-\mathbf{i}\arcsin\frac{\omega}{\Delta}}u\left(s_{R},s_{L}\right)\bar{u}\left(s_{L},s_{R}\right)-e^{\mathbf{i}\arcsin\frac{\omega}{\Delta}}\bar{u}\left(s_{R},s_{L}\right)u\left(s_{L},s_{R}\right)+2\mathbf{i}\sin\left(2\omega\left(s_{R}-s_{L}\right)+\varphi+\arcsin\dfrac{\omega}{\Delta}\right)}{2\cos\left(2\omega\left(s_{R}-s_{L}\right)+\varphi+2\arcsin\dfrac{\omega}{\Delta}\right)+\Tr\left\{ u\left(s_{R},s_{L}\right)\bar{u}\left(s_{L},s_{R}\right)\right\} }
\end{equation}
then $g\left(s_{0}\right)$ is obtained as \eqref{eq:g0} for $v_{x}>0$,
after injection of $g_{1}$ or $g_{2}$ in \eqref{eq:boundary}. One
gets
\begin{equation}
2g\left(s_{0}\right)\cos\arcsin\dfrac{\omega}{\Delta}=-2\mathbf{i}\sin\arcsin\dfrac{\omega}{\Delta}+\mathbf{i}u\left(s_{0},s_{L}\right)g_{1}u\left(s_{L},s_{0}\right)
\end{equation}
for the particle component (i.e. the component $11$ of the $\check{g}$
matrix) of eq.\eqref{eq:boundary} and one evaluates
\begin{equation}
g\left(s_{0}\right)=\dfrac{u\left(s_{0},s_{R}\right)\bar{u}\left(s_{R},s_{L}\right)u\left(s_{L},s_{0}\right)-u\left(s_{0},s_{L}\right)\bar{u}\left(s_{L},s_{R}\right)u\left(s_{R},s_{0}\right)-2\mathbf{i}\sin\left(2\omega\left(s_{R}-s_{L}\right)+\varphi+\arcsin\dfrac{\omega}{\Delta}\right)}{2\cos\left(2\omega\left(s_{R}-s_{L}\right)+\varphi+\arcsin\dfrac{\omega}{\Delta}\right)+\Tr\left\{ u\left(s_{R},s_{L}\right)\bar{u}\left(s_{L},s_{R}\right)\right\} }
\end{equation}
straightforwardly. The case $v_{x}<0$ is obtained by the solutions
in the superconducting electrodes 
\begin{align}
\check{g}\left(s\leq s_{L}\right) & =e^{\mathbf{i}\tau_{3}\varphi/4}\dfrac{e^{\mathbf{i}\eta/2}-\mathbf{i}\tau_{1}e^{\mathbf{i}\eta/2}}{\sqrt{2\cos\eta}}e^{-\tau_{3}\Delta\left(s-s_{L}\right)\tau_{3}\cos\eta}\left[g_{3}\tau_{-}-\tau_{3}\right]e^{\tau_{3}\Delta\left(s-s_{L}\right)\tau_{3}\cos\eta}\dfrac{e^{\mathbf{i}\eta/2}+\mathbf{i}\tau_{1}e^{-\mathbf{i}\eta/2}}{\sqrt{2\cos\eta}}e^{\mathbf{i}\tau_{3}\varphi/4}\nonumber \\
\check{g}\left(s\geq s_{R}\right) & =e^{-\mathbf{i}\tau_{3}\varphi/4}\dfrac{e^{\mathbf{i}\eta/2}-\mathbf{i}\tau_{1}e^{\mathbf{i}\eta/2}}{\sqrt{2\cos\eta}}e^{-\tau_{3}\Delta\left(s-s_{R}\right)\tau_{3}\cos\eta}\left[g_{4}\tau_{+}-\tau_{3}\right]e^{\tau_{3}\Delta\left(s-s_{L}\right)\tau_{3}\cos\eta}\dfrac{e^{\mathbf{i}\eta/2}+\mathbf{i}\tau_{1}e^{-\mathbf{i}\eta/2}}{\sqrt{2\cos\eta}}e^{-\mathbf{i}\tau_{3}\varphi/4}\label{eq:solution-supra-vneg}
\end{align}
instead of \eqref{eq:boundary}. It gives $\mathbf{Q}^{-1}\left[g_{4}\tau_{+}-\tau_{3}\right]=\left[g_{3}\tau_{-}-\tau_{3}\right]\mathbf{Q}^{-1}$
instead of \eqref{eq:continuity-LR}, with $\mathbf{Q}^{-1}\left(\varphi,s_{L},s_{R}\right)=\mathbf{Q}\left(-\varphi,s_{R},s_{L}\right)$.
One has thus $g_{4,3}\left(\varphi,s_{L},s_{R}\right)=g_{1,2}\left(-\varphi,s_{R},s_{L}\right)$.
Since $s=x/v_{x}$ one has, $s_{R}-s_{L}=L/v_{x}$ when $v_{x}>0$
and $s_{L}-s_{R}=L/\left|v_{x}\right|$ for $v_{x}<0$ when choosing
$s_{R,L}=\pm L/v_{x}$. One finally obtains \eqref{eq:g0} independent
of the sign of the velocity.

The $f_{0}$ matrix reads (we do not use this expression here, but
it is required to calculate perturbations, see e.g. \citep{Konschelle2015})
\begin{equation}
f\left(s\right)=-2\mathbf{i}\dfrac{e^{-2\mathbf{i}\omega\left(s_{L}-s\right)}e^{\mathbf{i}\frac{\varphi}{2}}e^{\mathbf{i}\arcsin\frac{\omega}{\Delta}}u\left(s,s_{R}\right)\bar{u}\left(s_{R},s\right)+e^{-2\mathbf{i}\omega\left(s_{R}-s\right)}e^{-\mathbf{i}\frac{\varphi}{2}}e^{-\mathbf{i}\arcsin\frac{\omega}{\Delta}}u\left(s,s_{L}\right)\bar{u}\left(s_{L},s\right)}{2\cos\left(2\omega\left(s_{R}-s_{L}\right)+\varphi+2\arcsin\dfrac{\omega}{\Delta}\right)+\Tr\left\{ u\left(s_{R},s_{L}\right)\bar{u}\left(s_{L},s_{R}\right)\right\} }\label{eq:f0}
\end{equation}
\end{widetext}at the point $s\in\left[s_{L},s_{R}\right]$. Eq.\eqref{eq:f0}
is given for positive velocity only $v_{x}>0$. The contribution $v_{x}<0$
is obtained by the substitution $\left(\varphi,s_{L},s_{R}\right)\rightarrow\left(-\varphi,s_{R},s_{L}\right)$
as before. One can calculate $\bar{f}\left(s\right)=\mathcal{T}f\left(s\right)\mathcal{T}^{-1}$
and $\bar{g}=\mathcal{T}g\mathcal{T}^{-1}$, the time-reversals of
$f$ and $g$, and then verify that $\check{g}^{2}=1$ straightforwardly.

\section{Short and long junction limits\label{app:Short-junction-limit}}

To understand how to get the short junction limit \eqref{eq:j-short-junction},
one writes
\begin{multline}
\sum_{\alpha,\beta=\pm}\alpha T_{\alpha\beta}\left(\omega\right)\\
=\sum_{\alpha,\beta=\pm}\tan\left(\dfrac{\varphi+\alpha\Phi}{2}+\beta\dfrac{\omega L}{\left|v_{x}\right|}+\beta\arcsin\dfrac{\omega}{\Delta}\right)\\
=\sum_{\alpha=\pm}\dfrac{2\sin\left(\varphi+\alpha\Phi\right)}{\cos\left(\varphi+\alpha\Phi\right)+\cos2\left(\dfrac{\omega L}{\left|v_{x}\right|}+\arcsin\dfrac{\omega}{\Delta}\right)}\label{eq:sum-alpha-T}
\end{multline}
with $T_{\alpha\beta}$ in \eqref{eq:T}, and playing with the parity
of the tangent, the sums over $\alpha$ and $\beta$ and finally using
the formula \eqref{eq:tangent} in order to isolate the terms in $\omega$.

The short junction verifies $\omega L/v_{F}\propto L/\xi_{T}\ll1$
and so one gets 
\begin{equation}
\lim_{L/\xi_{T}\ll1}\sum_{\alpha,\beta=\pm}\alpha T_{\alpha\beta}=\sum_{\alpha=\pm}\dfrac{2\sin\left(\varphi+\alpha\Phi\right)}{2\cos^{2}\dfrac{\varphi+\alpha\tau}{2}-\dfrac{\omega^{2}}{\Delta^{2}}}
\end{equation}
which can be converted to Matsubara frequencies $\omega=\mathbf{i}\omega_{n}=\mathbf{i}\pi k_{B}T\left(2n+1\right)$
and then sum over $n$. One obtains 
\begin{multline}
\sum_{n\geq0}\lim_{L/\xi_{T}\ll1}\sum_{\alpha,\beta=\pm}\alpha T_{\alpha\beta}=\\
\sum_{\alpha=\pm}\dfrac{\Delta}{2k_{B}T}\sin\dfrac{\varphi+\alpha\Phi}{2}\tanh\left(\dfrac{\Delta}{2k_{B}T}\cos\dfrac{\varphi+\alpha\Phi}{2}\right)
\end{multline}
using usual tricks to evaluate the sum
\begin{equation}
\sum_{n=0}^{\infty}\dfrac{1}{\alpha^{2}+\beta^{2}\left(n+1/2\right)^{2}}=\dfrac{\pi}{2\alpha\beta}\tanh\dfrac{\pi\alpha}{\beta}\;;\;\alpha,\beta\in\mathbb{R}
\end{equation}
see e.g. \citep{Mahan2000}. This is directly proportional to (the
sum over $\alpha$ of) $K_{\alpha}$ in \eqref{eq:K-short}.

For the spin observables, one uses the following tricks
\begin{multline}
\sum_{\alpha,\beta=\pm}\beta T_{\alpha\beta}\left(\omega\right)\\
=\sum_{\alpha,\beta=\pm}\tan\left(\dfrac{\alpha\varphi+\Phi}{2}+\beta\dfrac{\omega L}{\left|v_{x}\right|}+\beta\arcsin\dfrac{\omega}{\Delta}\right)\\
=\sum_{\alpha,\beta=\pm}\alpha\tan\left(\dfrac{\varphi+\alpha\Phi}{2}+\beta\dfrac{\omega L}{\left|v_{x}\right|}+\beta\arcsin\dfrac{\omega}{\Delta}\right)\label{eq:sum-beta-T}
\end{multline}
and so the sum is now odd in $\alpha$, which subsists in the expressions
for the spin density \eqref{eq:rho-spin-short} and the spin current
\eqref{eq:j-spin-short} in the short junction limit. The sum over
the Matsubara frequencies is the same as before and can be performed
irrespective of the presence of $\alpha$, hence one gets $\sum_{\alpha}\alpha K_{\alpha}$
in \eqref{eq:rho-spin-short} and \eqref{eq:j-spin-short}.

In the long junction limit, one starts again with either \eqref{eq:sum-alpha-T}
or \eqref{eq:sum-beta-T} but we apply this time $\omega L/v_{F}\propto L/\xi_{T}\gg1$
such that 
\begin{equation}
\lim_{L/\xi_{T}\gg1}\sum_{\alpha,\beta=\pm}\alpha T_{\alpha\beta}=\sum_{\alpha=\pm}\dfrac{2\sin\left(\varphi+\alpha\Phi\right)}{\cos\left(\varphi+\alpha\Phi\right)+\cos\dfrac{2\omega L}{\left|v_{x}\right|}}
\end{equation}
from \eqref{eq:sum-alpha-T}. Thus we have 
\begin{equation}
\lim_{L/\xi_{T}\gg1}\sum_{\alpha,\beta=\pm}\alpha T_{\alpha\beta}=2\sin\varphi\cos\Phi e^{-2\omega_{n}L/\left|v_{x}\right|}
\end{equation}
and so 
\begin{equation}
\sum_{n\geq0}\lim_{L/\xi_{T}\gg1}\sum_{\alpha,\beta=\pm}\alpha T_{\alpha\beta}=\dfrac{\sin\varphi\cos\Phi}{\sinh\dfrac{\pi TL}{\left|v_{x}\right|}}
\end{equation}
\begin{equation}
\sum_{n\geq0}\lim_{L/\xi_{T}\gg1}\sum_{\alpha,\beta=\pm}\beta T_{\alpha\beta}=\dfrac{\cos\varphi\sin\Phi}{\sinh\dfrac{\pi TL}{\left|v_{x}\right|}}
\end{equation}
since the sum over the Matsubara frequencies can be performed easily
\begin{equation}
\sum_{n=0}^{\infty}e^{-\alpha\left(n+1/2\right)}=\dfrac{1}{2\sinh\dfrac{\alpha}{2}}\;;\;\alpha\in\mathbb{R}_{+}
\end{equation}
as a geometric progression. In the long junction limit, the trajectories
$v_{x}$ with large angles from the junction axis are killed exponentially
(say the trajectories $v_{x}=v_{F}\cos\phi$ with $\phi\approx\pi/2$
for a circular Fermi surface) and do not participate to the transport.
In the long junction limit, the Andreev bound states are equally spaced,
and we recover the effective action of a harmonic oscillator.

\end{document}